\documentclass[%
showpacs,
amsmath,amssymb,
aps,
pra,
floatfix,
twocolumn,
longbibliography
]{revtex4-1}
\usepackage{graphicx}
\usepackage{dcolumn}
\usepackage{bm}
\usepackage{color}
\usepackage{comment}
\usepackage{diagbox}
\usepackage{verbatim}
\usepackage{multirow}

\begin{document}


\title{Quantum annealing in SU(3) multiplet space with nonlocal drivers}

\author{Yang Wei Koh}
\email{patrickkyw@gmail.com}

\affiliation{}

\date{\today}

\begin{abstract}

A theoretical framework for quantum annealing based on $\mathfrak{su}(3)$ algebra is proposed, and applied to the problem of overcoming first-order transitions in rugged energy landscapes. Conservation of the Casimir invariant means that one can work with irreducible representations of $\mathfrak{su}(3)$, avoiding the exponentially large Hilbert spaces of spin glass systems. In this framework, quantum drivers exhibit nonlocal properties in the sense that during annealing the wave function can be transported far away from a local minimum, thereby avoiding being trapped by it. We consider Hamiltonians with two quantum drivers and studied them numerically. It is shown that energy gap closures can be circumvented via a suitable path in the parameter space of the two drivers. Comparison with more traditional annealing driven by transverse field and antiferromagnetic operators suggests that $\mathfrak{su}(3)$ drivers are more effective in attaining the global minimum of rugged energy landscapes.

\end{abstract}
\pacs{}
\maketitle



\section{Introduction}
\label{sec.introduction}

Quantum annealing (QA) \cite{ Kadowaki98, Farhi01, Das08, Das15, Albash18, Hauke20} has emerged as one of the central focus in quantum computation in recent years. Also known as adiabatic quantum computation, it is a metaheuristic for solving optimization problems via careful control of quantum fluctuations. In the last decade, immense interest in this field has been driven by significant advancements in D-Wave hardware, notably in the optimization of spin glasses \cite{Venturelli15}, tunneling \cite{Boixo16,Denchev16}, quantum simulations \cite{Harris18, Nishimura20, Bando20, King23,King25}, and continuous variable minimization \cite{Abel21,Arai23}, to name a few. On the theoretical side, many fundamental aspects of QA are also now well-understood, such as concerning the convergence of QA \cite{Morita08,Kimura22,Kimura23}, the role of adiabaticity \cite{Suzuki05,Takahashi17, Takahashi19} and energy gap \cite{Jorg10, Bapst12, Seki12, Seoane12, Okuyama15}, and speedups via nonlinear schedules \cite{Morita07, Matsuura21, Susa21, Koh22} and nonstoquastic drivers \cite{Seki12, Seoane12, Seki15, Nishimori17, Durkin19, Albash19, Takada20, Koh20, Takada21}. Among these, the issue of energy gap closure and the accompanying problem of first-order phase transition is one of the important challenges currently faced by QA. Broadly speaking, the probability that the ground state undergoes transition to the first-excited state depends on the minimum energy gap between them during the course of annealing. A small gap means that the bottleneck region must be traversed slowly to avoid a transition, resulting in a long annealing time. The crucial question is how the gap scales with system size. If it decreases exponentially with size, one is faced with a first-order phase transition whereby the wave function must make a macroscopic jump in configuration space in order to remain in the ground state, which is a very difficult obstacle to overcome. This predicament has been documented extensively both in non-frustrated \cite{Jorg10, Bapst12, Seki12, Seoane12, Durkin19, Albash19, Koh20} as well as frustrated systems \cite{Seki15, Takada21, Jorg08, Young08, Young10, Knysh16, Koh18}. 

In non-frustrated systems, the annealing Hamiltonians are frequently defined in terms of angular momentum operators
\begin{equation}
J_{\alpha}=\sum_{i=1}^N\sigma_i^{\alpha}
\label{eq.Jxyz.definition}
\end{equation}
where $\alpha$ $(=x,y,z)$ labels the Cartesian axes, $\sigma_i^{\alpha}$ is the $\alpha$-directional Pauli matrix of the $i$th spin, and $N$ is the total number of spins. These are usually fully-connected ferromagnetic or antiferromagnetic systems whose Hamiltonians commute with the total angular momentum operator $J^2=J_x^2+J_y^2+J_z^2$. The advantage of such systems is that they are amenable to numerical studies when the system size $N$ is large, since one can just focus on an irreducible subspace, avoiding an otherwise exponentially large Hilbert space. However, the limitation is that their energy landscapes are usually smooth and rather simple, quite unlike the rugged and complex energy surfaces that one encounters in realistic optimization problems. On the other hand, the QA of frustrated systems has also received much attention \cite{Seki15, Takada21, Jorg08, Young08, Young10, Knysh16, Koh18, Liu15, Mukherjee18}. As their Hilbert spaces are exponentially large, QA simulations of such systems are usually limited to small system sizes ($N \lesssim 30$) \cite{Jorg08,Takada21}, or performed using computationally expensive quantum Monte Carlo methods \cite{Young08,Young10,Liu15,Mukherjee18}. In view of the above difficulties, there are ongoing efforts to develop new theoretical approaches for studying quantum spin glasses \cite{Smelyanskiy25} and their QA \cite{Knysh16}. 

\begin{figure}[h]
\begin{center}
\includegraphics[scale=0.4]{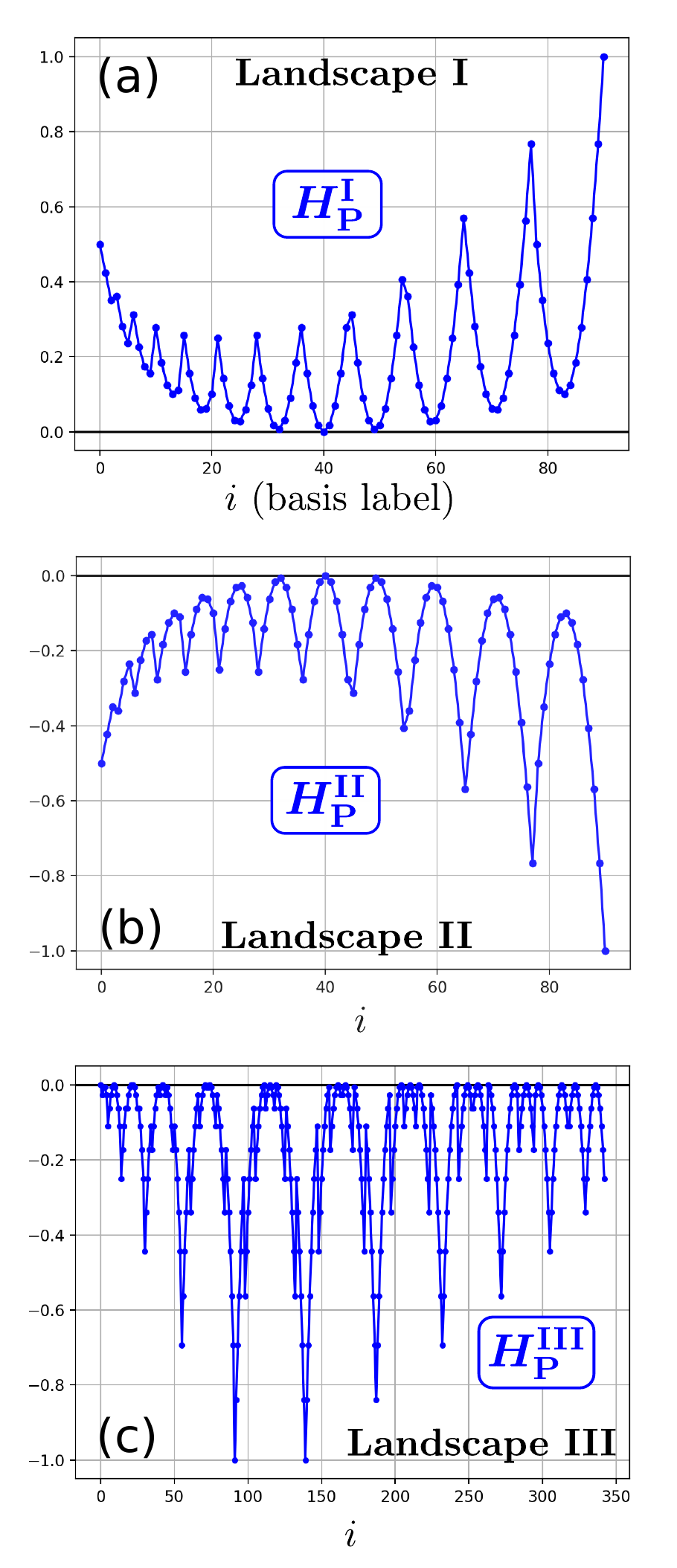}
\caption{Energy landscapes studied in this work. Panels (a) to (c) show the diagonal matrix elements of problem Hamiltonians $H_{\mathrm{P}}^{\mathrm{I}}$ to $H_{\mathrm{P}}^{\mathrm{III}}$, defined in the text. Horizontal axis labels the basis vectors of the irreducible representation of the $\mathfrak{su}(3)$ multiplet, following Refs. \cite{Shurtleff23,Shurtleff24}. Vertical axis has been rescaled such that the most positive (or negative) matrix element has unit magnitude. Connecting lines are to guide the eye only.}
\label{fig.4 su3 potentials}
\end{center}
\end{figure}

In this paper, we propose a theoretical framework to study QA in scenarios which are intermediate between non-frustrated systems and full-blown spin glasses, with a particular focus on the problem of overcoming energy gap closures in systems with demonstrably rugged energy landscapes. We believe that there are two issues to be addressed when approaching this fundamental problem. The first concerns how to set up a non-trivial system whose Hilbert space is manageable, while at the same time possessing a landscape that is rugged yet controllable. The latter (often overlooked) quality is important for us to be able to infer and identify underlying physical mechanisms from simulation data. Traditional spin glass systems, such as the Sherrington-Kirkpatrick model, have exponentially large Hilbert spaces whose energy landscapes are unpredictable and highly-dependent on specific instances. Our idea is to consider a larger Lie algebra than the angular momentum algebra. The Pauli matrices in Eq. (\ref{eq.Jxyz.definition}) arise as generators of the group of special unitary transformations in two dimensions, also known as $\mathfrak{su}(2)$ \cite{footnote.su.convention}. We propose using the elements of $\mathfrak{su}(3)$ \cite{Gasiorowicz66, Pfeifer03, Georgi19} to construct our annealing Hamiltonians. Consider generalizing Eq. (\ref{eq.Jxyz.definition}) as
\begin{equation}
\Lambda_{\alpha}=\sum_{i=1}^N\lambda_i^{\alpha}
\label{eq.Gell-mann.aggregates}
\end{equation}
where $\lambda_i^{\alpha}$ is the $\alpha$th Gell-Mann matrix of the $i$th particle, and the integer label $\alpha\in\{1,\cdots,8\}$. Gell-Mann matrices are generators of $\mathfrak{su}(3)$ in its fundamental representation \cite{Gasiorowicz66, Pfeifer03}. As mentioned earlier, Hamiltonians defined in terms of $J_{\alpha}$ commute with $J^2$, so one works with the spin-$j$ multiplet belonging to $J^2=j(j+1)$, which is $(2j+1)$-dimensional. Similarly in $\mathfrak{su}(3)$, annealing Hamiltonians which are built from $\Lambda_{\alpha}$ commute with the $\mathfrak{su}(3)$ quadratic Casimir operator [see Eq. (\ref{eq.Casimir.C1.definition})], so we can work within the irreducible representation of a $\mathfrak{su}(3)$ multiplet, avoiding an exponentially large Hilbert space. In $\mathfrak{su}(3)$, a multiplet is labeled by a pair of non-negative integers $(p,q)$, and its dimension is \cite{Gasiorowicz66,Pfeifer03}
\begin{equation}
d_m(p,q)=\frac{1}{2}(p+1)(q+1)(p+q+2)
\label{eq.dm(p,q).formula}
\end{equation}
The multiplet dimension of $\mathfrak{su}(3)$ therefore increases faster than $\mathfrak{su}(2)$, but slower than that of a full-blown spin glass. In return for this increase in dimensionality, $\mathfrak{su}(3)$ allows us to design a wider variety of annealing models compared to $\mathfrak{su}(2)$. An irreducible representation of $\mathfrak{su}(3)$ has two diagonal generators which are linearly independent of each other. The two commuting generators, known as the Cartan subalgebra, give us more freedom in terms of sculpting the energy landscape that one wishes to optimize. [In comparison, $\mathfrak{su}(2)$ possesses only one such generator, usually taken to be the $J_z$ operator.] Figure \ref{fig.4 su3 potentials} shows the three energy landscapes (i.e. problem Hamiltonians) that we will be studying in this paper. Details will be presented later. Here, we would just like to point out that such rugged surfaces are not easily achievable using $J_z$ within the $\mathfrak{su}(2)$ framework. These landscapes capture the most salient characteristic of spin glasses, namely the many-valleyed structure of a glassy energy surface. Although not necessarily as complex as a full-blown spin glass, we believe that studying such landscapes within the manageable Hilbert space of a $\mathfrak{su}(3)$ multiplet can aid in understanding the fundamental aspects of QA in glassy scenarios. 

In Eq. (\ref{eq.Jxyz.definition}), the Pauli matrices act on two-component qubits, whereas the Gell-Mann matrices in Eq. (\ref{eq.Gell-mann.aggregates}) act on three-component states, or qutrits, which are currently receiving much attention in diverse areas of quantum computation \cite{Roy23,Goss24,Bottrill23}. Traditionally in $\mathfrak{su}(3)$, such states arise in elementary particle physics as sakaton or quarks \cite{Gasiorowicz66,Georgi19}. More pertinent to our work is perhaps the bifurcation-based approach to adiabatic quantum computation proposed by Goto \cite{Goto16a, Goto16b, Goto19}, in which a qutrit-like unit is represented by a Kerr nonlinear oscillator. Takahashi reformulated the oscillator as a spin-1 particle, and demonstrated some advantages of bifurcation-based QA vis-\`a-vis more traditional QA utilizing qubits \cite{Takahashi22}. A much clearer insight into the relation between oscillator and spin based QA was subsequently given by Miyazaki \cite{Miyazaki22}, who mapped the two approaches and showed how the spin-1 feature arise from the bosonic operators. In these works, much emphasis has been placed on the three-component and bifurcational aspects of the oscillator formulation. Although interesting in its own right, we think that an even richer and more prospective theoretical framework can be achieved if we look at things from the Lie algebra perspective.

The second issue concerns how to overcome first-order transitions in QA. In this respect, the use of so-called nonstoquastic drivers has been shown to be quite successful \cite{Seki12, Seoane12, Seki15, Nishimori17, Durkin19, Albash19, Takada20, Koh20, Takada21}. A distinguishing feature of QA is the flexibility to select appropriate kinetic drivers. In $\mathfrak{su}(2)$, there are three generators and usually $J_x$, in the form of a transverse field, is used as the driver. When faced with a first-order transition, it has been shown that adding a second driver in the form of antiferromagnetic interaction $+(J_x)^k$ ($k\in\mathbb{Z}^+$) can soften the transition into a second-order one \cite{Seki12, Seoane12, Durkin19, Koh20}. In the case of $\mathfrak{su}(3)$, there are eight generators, so (excluding the Cartan subalgebra) we have six linearly independent drivers to choose from, which can be employed individually or in pairs. The novel feature of $\mathfrak{su}(3)$ drivers is that the irreducible representation of their matrix operators exhibit nonlocal properties, in the sense that they possess matrix elements that connect distant states directly. During annealing, this property offers the wave function a way of escaping a local minimum, which enables us to overcome the problem of first-order transitions. By contrast, the transverse field and antiferromagnetic drivers are comparatively more local in that they connect only the neighboring states in Hilbert space. Indeed, we will demonstrate that $\mathfrak{su}(3)$ drivers can be very effective in situations where the transverse field driver fails. We would also like to point out that unlike antiferromagnetic drivers, the primary $\mathfrak{su}(3)$ drivers are not polynomial powers of each other. In other words, the $\mathfrak{su}(3)$ framework employs only the two-driver feature of QA, and may not need to involve nonstoquasticity, which makes quantum Monte Carlo simulations challenging to implement \cite{Ohzeki17}. Lastly, two-driver annealing have hitherto been applied on $J_{\alpha}$-systems \cite{Seki12, Seoane12, Albash19, Durkin19, Koh20}, a variety of locally-connected models \cite{Albash19, Takada21}, and mildly frustrated systems \cite{Seki15}. Their effectiveness in demonstrably rough energy landscapes has not received much attention, to the best of our knowledge. Overall, we believe that the $\mathfrak{su}(3)$ framework offers more variety in terms of the types of energy landscapes and the combinations of quantum drivers that can be studied in QA.

The rest the paper is organized as follows. Section \ref{sec.su2 and3 review} presents a brief review of $\mathfrak{su}(3)$ algebra and its multiplet structure. Section \ref{sec.quantum drivers} discusses $\mathfrak{su}(3)$ quantum drivers, elaborating upon our earlier comments that they possess nonlocal properties. Sections \ref{sec.landscape I}, \ref{sec.landscape II}, and \ref{sec.landscape III.pure} present the results of numerical simulations and form the main body of the paper. Each section focuses on one landscape in Fig. \ref{fig.4 su3 potentials}. Both static (energy gap and ground state eigenfunction) as well as annealing (residual energy and wave function dynamics) behaviors are examined in detail. We study Hamiltonians with two quantum drivers, and demonstrate how one can choose annealing paths that avoids energy gap closures and overcome first-order transitions. Particular attention is also being paid to dynamics of the wave function in $\mathfrak{su}(3)$ multiplet space, illustrating how the multiplet structure helps in the attainment of global minimum. As comparison, we also carried out annealing using more familiar drivers, to offer some perspectives on how our $\mathfrak{su}(3)$ models perform relative to traditional QA. Section \ref{sec.Jx only} studies the three landscapes driven by the transverse field operator $J_x$ alone. In Sec. \ref{sec.Jx and another}, we try improving upon $J_x$ by combining it with various secondary drivers (e.g., antiferromagnetic interaction). Through these comparative studies, we hope to demonstrate the advantages of $\mathfrak{su}(3)$ algebra as a framework for QA of systems with rugged energy landscapes. Finally, Sec. \ref{sec.conclusions} summarizes and concludes the paper. Some technical details on $\mathfrak{su}(3)$ are provided in Appendix \ref{app.sec.irrep of su(3)}.


\section{SU(3) and its multiplet structure}
\label{sec.su2 and3 review}

The Lie algebra $\mathfrak{su}(3)$ arises as generators of the group of special unitary transformations in three dimensions. Instead of $\Lambda_{\alpha}$ in Eq. (\ref{eq.Gell-mann.aggregates}), we follow the so-called \emph{TUV} notation and denote the eight real-valued generators as $T_3, U_3, T_{\pm},U_{\pm},$ and $V_{\pm}$. They satisfy the commutator relations \cite{Gasiorowicz66, Pfeifer03, Georgi19}
\begin{gather}
[T_3,U_3]=0
\label{eq.commutator.Cartan}
\\[3pt]
\begin{array}{cc}
[T_3,T_{\pm}]=\pm T_{\pm}\,,
& \,\, [T_+,T_-]=2T_3
\end{array}
\label{eq.commutator.subalgebra.T}
\\[3pt]
\begin{array}{cc}
[U_3,U_{\pm}]=\pm U_{\pm}\,,
& \,\, [U_+,U_-]=2U_3
\end{array}
\label{eq.commutator.subalgebra.U}
\\[3pt]
[V_+,V_-]=2(T_3+U_3)
\label{eq.commutator.subalgebra.V}
\\[3pt]
\begin{array}{cc}
       [T_3,U_{\pm}]=\mp\frac{1}{2}U_{\pm}\,,
& \,\, [T_3,V_{\pm}]=\pm\frac{1}{2}V_{\pm}\\
\end{array}
\label{eq.commutator.eigen.T3}
\\[3pt]
\begin{array}{cc}
       [U_3,T_{\pm}]=\mp\frac{1}{2}T_{\pm}\,,
& \,\, [U_3,V_{\pm}]=\pm\frac{1}{2}V_{\pm}\\
\end{array}
\label{eq.commutator.eigen.U3}
\\[3pt]
[T_{\pm},V_{\pm}]=[U_{\pm},V_{\pm}]=[T_{\pm},U_{\mp}]=0
\label{eq.commutators.cross.zero}
\\[3pt]
\begin{array}{ccc}
[T_{\pm},U_{\pm}]=\pm V_{\pm}\,,  
& \,\, [T_{\pm},V_{\mp}]=\mp U_{\mp}\,, 
& \,\, [U_{\pm},V_{\mp}]=\pm T_{\mp}
\end{array}
\label{eq.commutators.cross.non-zero}
\end{gather}
The commuting generators $T_3$ and $U_3$ span the Cartan subalgebra, and are used to construct our energy landscapes. One can define another element $V_3=T_3+U_3$ associated with $V_{\pm}$. Then, Eqs. (\ref{eq.commutator.subalgebra.T}) to (\ref{eq.commutator.subalgebra.V}) show that $\mathfrak{su}(3)$ contains three $\mathfrak{su}(2)$ subalgebras $\{T_3,T_{\pm}\}$, $\{U_3,U_{\pm}\}$, and $\{V_3,V_{\pm}\}$. The generators $T_{\pm},U_{\pm},$ and $V_{\pm}$ act as ladder operators for the eigenvalues of $T_3$ and $U_3$. They are analogous to $J_{\pm}$ of $\mathfrak{su}(2)$ and will serve as quantum drivers.

The eight generators commute with the quadratic Casimir invariant \cite{Pfeifer03}
\begin{equation}
C_1=T_-T_+ + V_-V_+ + U_+U_-
+ T_3 + V_3 - U_3
+ T_3^2 + \frac{1}{3}(V_3+U_3)^2
\label{eq.Casimir.C1.definition}
\end{equation}
Hence, a Hamiltonian defined in terms of $\mathfrak{su}(3)$ generators commutes with $C_1$, so one only needs to work with irreducible representations of $\mathfrak{su}(3)$. This is similar to $\mathfrak{su}(2)$ where $J^2$ is conserved.

\begin{figure}
\begin{center}
\includegraphics[scale=0.45]{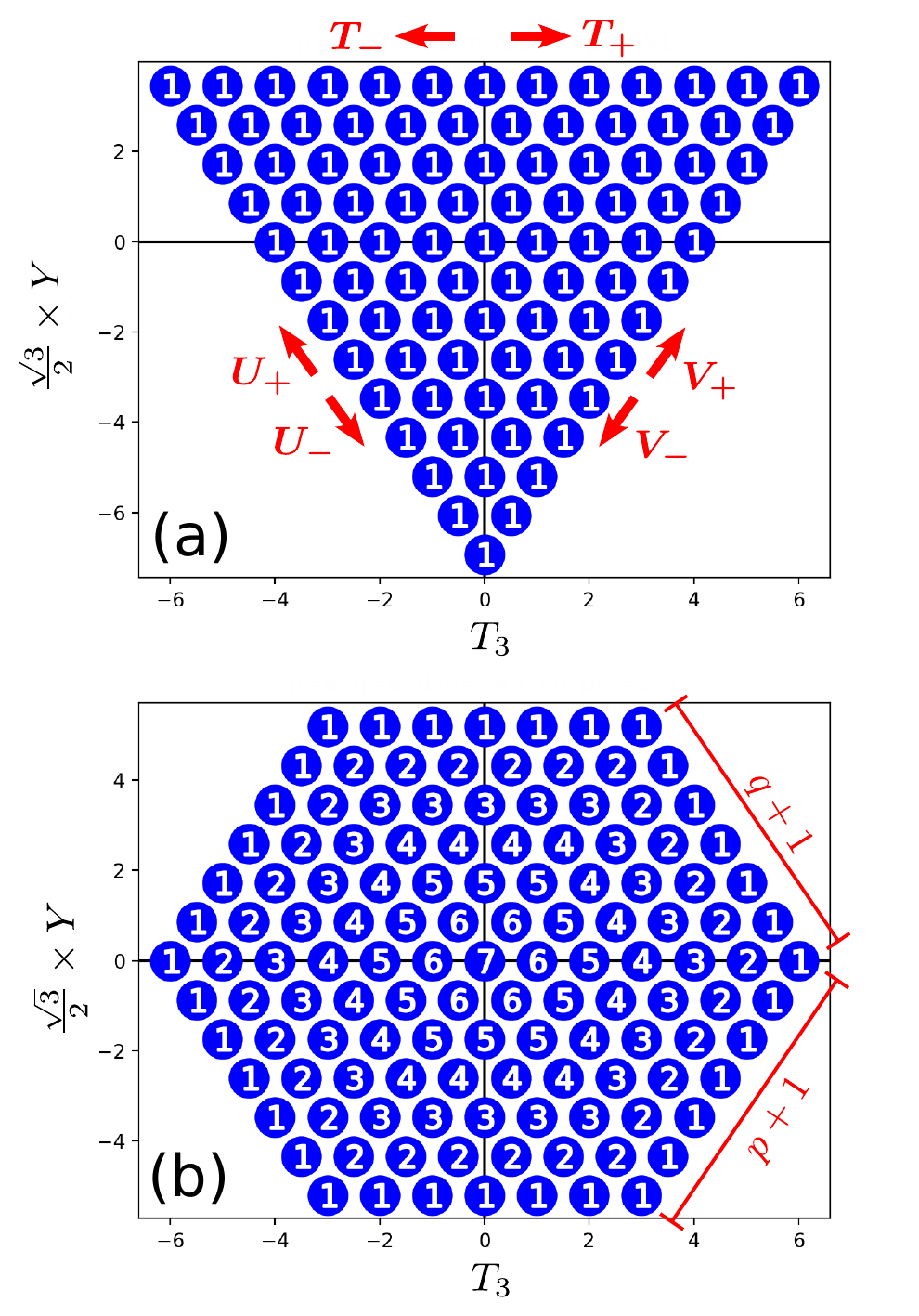}
\caption{The $T_3$ versus $Y$ plots of $\mathfrak{su}(3)$ multiplets. (a) For $(p,q)=(12,0)$. (b) For $(6,6)$. The states in a multiplet are arranged in the form of a hexagon, which can be asymmetric in general. Panel (b) shows the geometric significance of $p+1$ and $q+1$ as the lengths of two of its sides. The integer within each circle indicates the multiplicity at the point $(T_3,Y)$. Arrows in panel (a) indicate the directions in which states are translated within the multiplet when acted upon by the ladder operators $T_{\pm}, U_{\pm}$, and $V_{\pm}$.}
\label{fig.su3 YvsT3 plot}
\end{center}
\end{figure}

A $\mathfrak{su}(3)$ multiplet is characterized by a pair of non-negative integers $(p,q)$, containing $d_m(p,q)$ states [Eq. (\ref{eq.dm(p,q).formula})]. Geometrically, the states are arranged in the form of an asymmetric hexagon with $p+1$ and $q+1$ being the lengths of two of its sides. Figure \ref{fig.su3 YvsT3 plot} shows the multiplet diagrams for $(p,q)=(12,0)$ and $(6,6)$. The geometrical significance of $p$ and $q$ is indicated in panel (b). When $q$ (or $p$) is zero, the hexagon becomes an equilateral triangle [panel (a)]; when $p=q$, a regular hexagon is obtained [panel (b)].

Each state in a multiplet is labeled by two quantum numbers, the eigenvalues of $T_3$ and the hypercharge operator $Y=\frac{2}{3}(T_3+2U_3)$, which are plotted along the horizontal and vertical axes of the multiplet diagram. More than one state can have the same set of quantum numbers, so each point $(T_3,Y)$ in the diagram is characterized by an integer called multiplicity, which indicates the number of states at that point. In Fig. \ref{fig.su3 YvsT3 plot}, the integer in each circle indicates the multiplicity at $(T_3,Y)$. The sum of all multiplicities gives $d_m(p,q)$ \cite{Pfeifer03}.

In this paper, we work with the irreducible representations of $\mathfrak{su}(3)$. These are matrix forms of the generators represented using the states of a $\mathfrak{su}(3)$ multiplet as basis. Recent contributions by Shurtleff provide numerical programs to let the user call these matrices \cite{Shurtleff23,Shurtleff24}. In the following, we simply use the matrices obtained from the algorithm. Some details on how the states in a multiplet are labeled are given in Appendix \ref{app.sec.irrep of su(3)}.


\section{Nonlocal quantum drivers}
\label{sec.quantum drivers}

\begin{figure}[h]
\begin{center}
\includegraphics[scale=0.4]{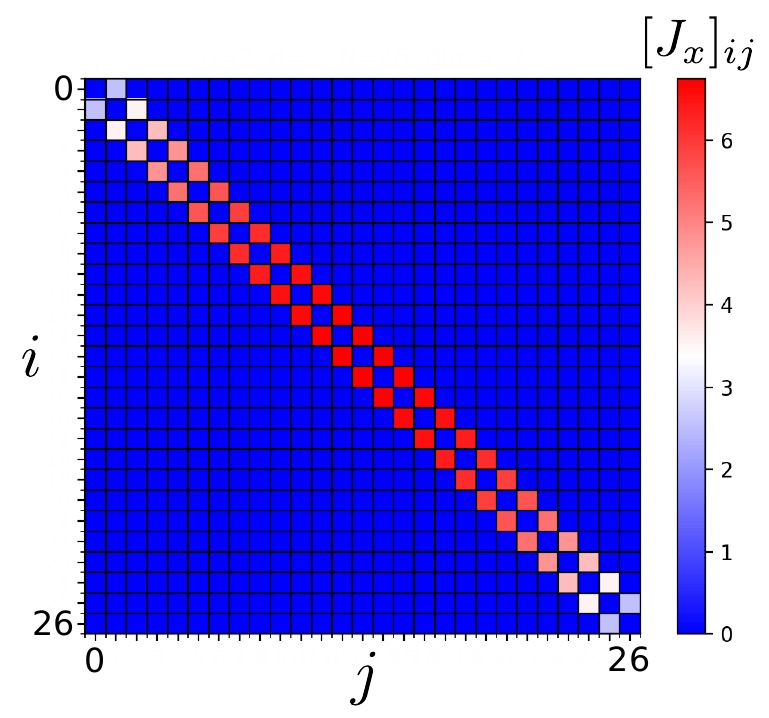}
\caption{Irreducible representation of the angular momentum matrix $J_x$, plotted as a heat map with the color of each cell indicating the value of the corresponding matrix element. The spin $j=13$ multiplet is shown ($\hbar=1$). The tridiagonal form signifies the local nature of the driver.}
\label{fig.Jx and Jx2.heatmap}
\end{center}
\end{figure}

Before discussing annealing, let us briefly compare $\mathfrak{su}(3)$ drivers with traditional ones, to highlight the novel features of the $\mathfrak{su}(3)$ framework. Figure \ref{fig.Jx and Jx2.heatmap} shows the standard irreducible representation of the angular momentum matrix $J_x$ for the spin $j=13$ multiplet. The matrix is presented as a heat map, where the color at the $i$th row $j$th column indicates the numerical value of the matrix element $[J_x]_{ij}$. One sees that $J_x$ is a tridiagonal matrix, and it is in this sense that we say it is a local operator. The matrix connects only neighboring states, so the probability amplitude of the wave function at state index $i$ can only be transported locally (to states $i-1$ and $i+1$) during time evolution. Another perspective is to note that the tridiagonal form of $J_x$ is similar to the matrix representation of the differential operator $\frac{d}{dx}$ under the finite difference scheme. The momentum operator $\frac{\hbar}{i}\frac{d}{dx}$ performs spatial translation along the $x$-direction by a small amount, i.e. locally.

\begin{figure*}
\begin{center}
\includegraphics[scale=0.3]{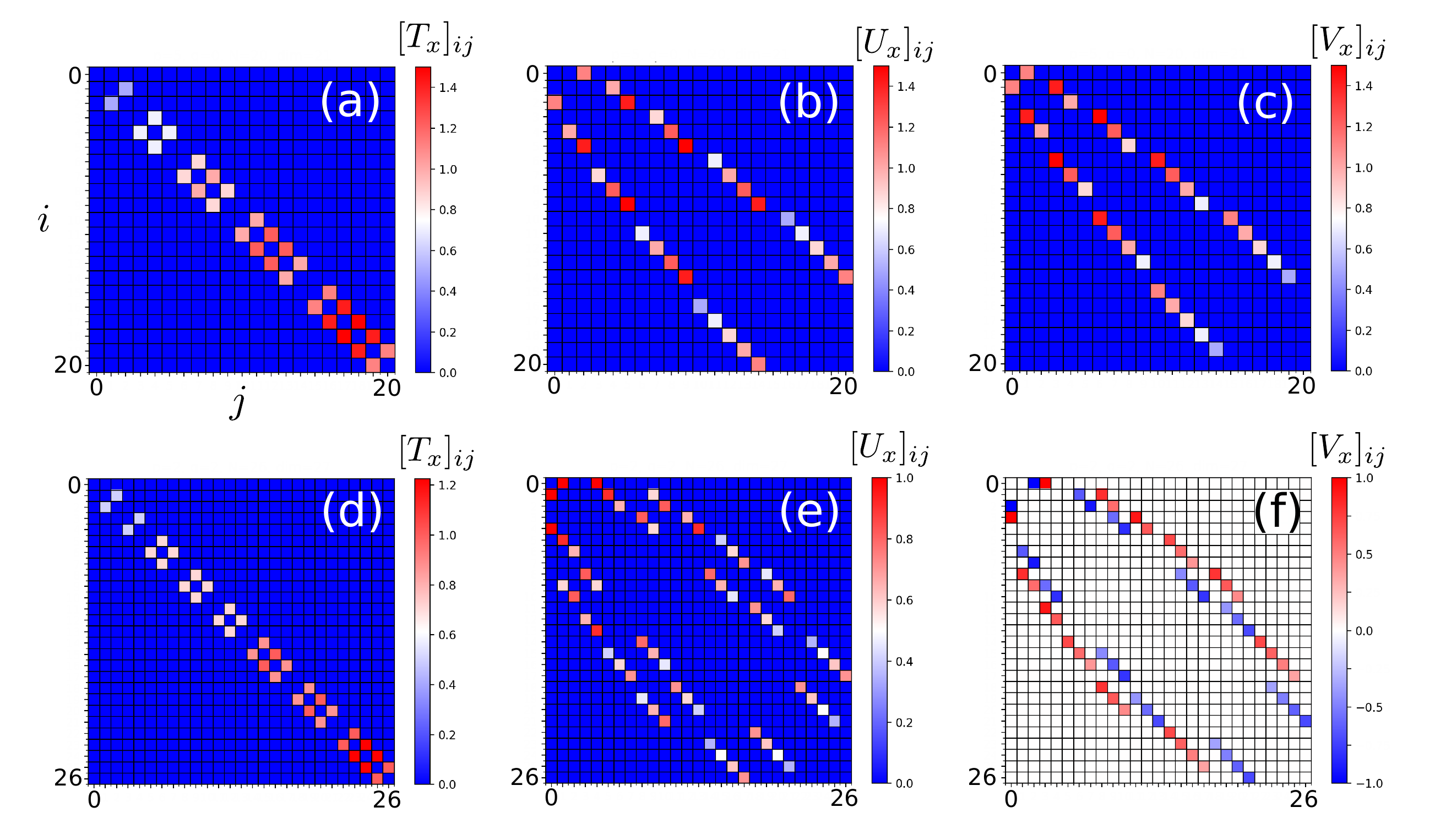}
\caption{Quantum drivers of $\mathfrak{su}(3)$. The matrices of $T_x$ (left column), $U_x$ (center), and $V_x$ (right) are plotted in the manner of Fig. \ref{fig.Jx and Jx2.heatmap}. Upper row shows the drivers of the $(p,q)=(5,0)$ multiplet, lower row those of the $(2,2)$ one. One sees that the drivers $U_x$ and $V_x$ have non-zero matrix elements far away from the main diagonal. These off-diagonal elements connect distant locations in configuration space directly, bringing about nonlocal interactions.}
\label{fig.su3 drivers.heatmap}
\end{center}
\end{figure*}

For $\mathfrak{su}(3)$, define the operator
\begin{equation}
T_x=\frac{1}{2}\left(T_+ + T_-\right)
\label{eq.Tx.Ux.Vx.definitions}
\end{equation}
and similarly for $U_{x}$ and $V_{x}$. Figure \ref{fig.su3 drivers.heatmap} shows some examples of their irreducible representations, obtained from Shurtleff's algorithm. The matrices are presented as heat maps as in Fig. \ref{fig.Jx and Jx2.heatmap}. The upper row shows the $(5,0)$-multiplet representation, the lower row the $(2,2)$ one. The left, center, and right columns show $T_x$, $U_x$, and $V_x$, respectively. We see that $T_x$ is tridiagonal and therefore local. On the other hand, $U_x$ and $V_x$ have non-zero matrix elements far from the main diagonal. This means that states which are far apart in configuration space can be connected directly by these operators. We use the term nonlocal in this sense. Nonlocality is useful for QA because when the system is stuck in a local minimum, it offers a way for the wave function to `leak' to a distant location and avoid being trapped. Note that nonlocal moves are by no means unfamiliar to computational physics. For instance, in replica exchange Monte Carlo one performs swaps between macroscopically different state configurations to help spin glass systems relax \cite{Hukushima96}. Another example is the Swendsen-Wang cluster algorithm for sampling systems near criticality \cite{Swendsen87}.

The nonlocal connections in the matrix elements of $U_x$ and $V_x$ have another interpretation as translations in the $\mathfrak{su}(3)$ multiplet diagram. In Fig. \ref{fig.su3 YvsT3 plot}(a), the red arrows indicate the directions in which states are being translated when acted upon by the ladder operators $T_{\pm}$, $U_{\pm}$, and $V_{\pm}$. As mentioned earlier, $\mathfrak{su}(3)$ has three $\mathfrak{su}(2)$ subalgebras.  Each subalgebra translates along the direction of its own $\mathfrak{su}(2)$ multiplet. For example, the multiplets of the $T$-subalgebra are aligned horizontally, so $T_{\pm}$ translate along the horizontal direction. For the $U$- and $V$-subalgebras, their multiplets are tilted at $120^{\circ}$ and $60^{\circ}$ polar angles. When viewed in multiplet space, what appears as nonlocal moves in configuration space are actually local, since $U_{\pm}$ and $V_{\pm}$ are merely translating to the nearest neighboring state within their own multiplet.


\section{Landscape I: Convex basin with rugged interior}
\label{sec.landscape I}

\subsection{Problem Hamiltonian}
\label{}

As our first system, we look at Landscape I. The problem Hamiltonian is defined in terms of irreducible representations of $T_3$ and $U_3$ as
\begin{equation}
H_{\mathrm{P}}^{\mathrm{I}}=\frac{1}{\tilde{n}}
\left(
T_3^2+U_3^2
\right)
\label{}
\end{equation}
The factor $\tilde{n}=\sqrt{d_m-1}$ ensures a proper scaling of energy with system size. Figure \ref{fig.4 su3 potentials}(a) shows the graph of $[H_{\mathrm{P}}^{\mathrm{I}}]_{ii}$ versus $i$ for the $(12,0)$-multiplet. Landscape I simulates a bowl with rugged interior. Such a scenario occurs, for example, in the quantum Hopfield model \cite{Nishimori96}, where to retrieve a stored pattern one must attain the global minimum of one of many memory basins. It has been found that these basins have rugged interiors \cite{Knysh16,Koh18}. This landscape mimics the situation whereby one is inside one of the memory basins and must search for the lowest minimum of that basin.

\subsection{Circumventing annealing bottlenecks on gap landscape}
\label{subsec.Landscape I.statics}

\begin{figure}[h]
\begin{center}
\includegraphics[scale=0.35]{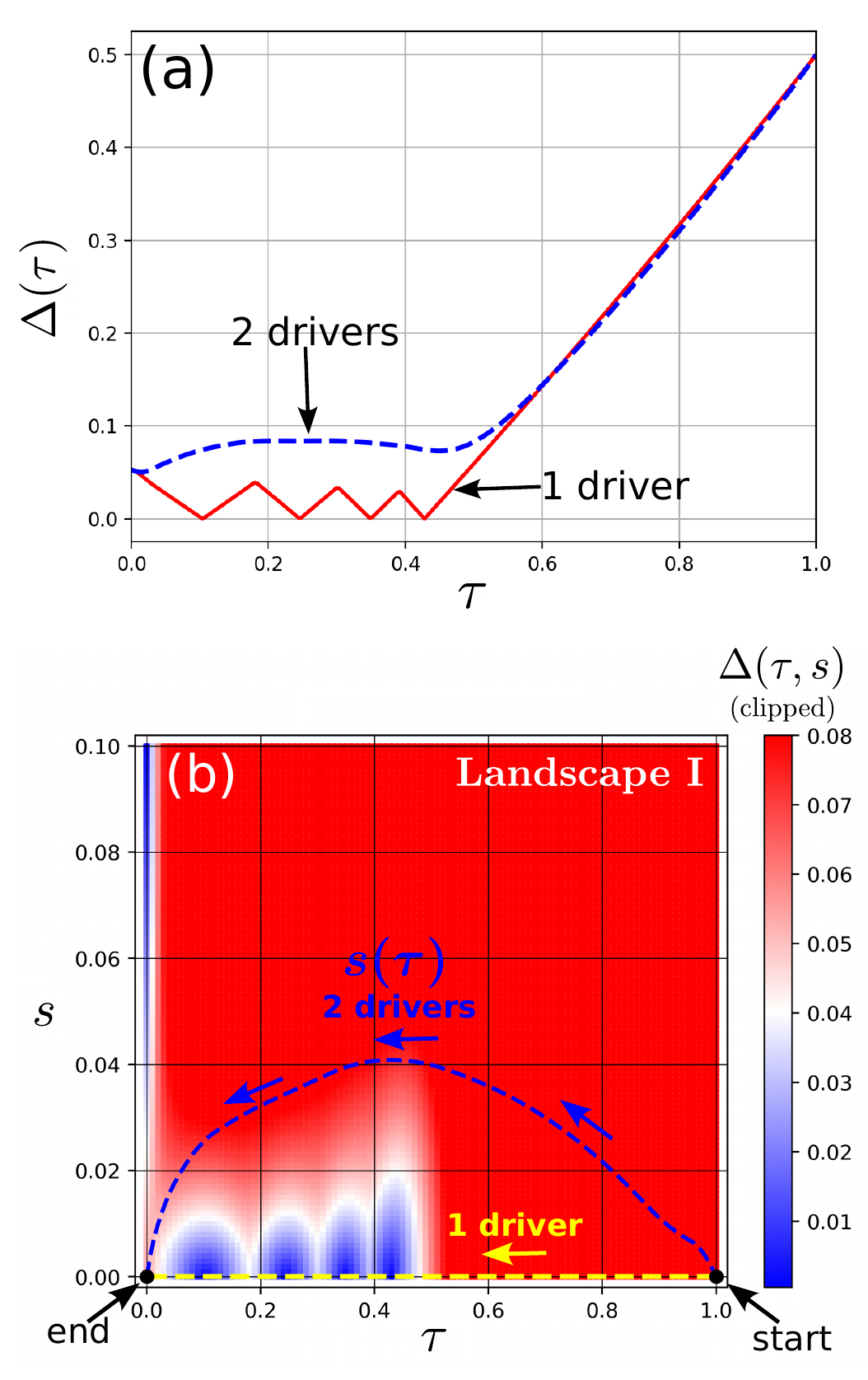}
\caption{Energy gaps of Landscape I. (a) Solid line (red) shows $\Delta(\tau)$ of the 1-driver $H^{(1)}$. As $\tau$ decreases, there are four gap closures. Dashed line (blue) shows that of 2-driver $H^{(2)}$, evaluated along the path $s(\tau)$ shown in panel (b). There are no closures along this path. (b) Gap landscape $\Delta(\tau,s)$ of $H^{(2)}(\tau,s)$, plotted as a heat map. Cool colors indicate small gaps. To accentuate the small-gap region, the value of $\Delta(\tau,s)$ is clipped at 0.08. Dashed curves show the annealing paths. The 1-driver path (yellow) traverses the small-gap region, while the 2-driver path (blue) avoids it. Arrows indicate the direction in which annealing proceeds.}
\label{fig.gaps.Landscape I}
\end{center}
\end{figure}

\begin{figure}[h]
\begin{center}
\includegraphics[scale=0.6]{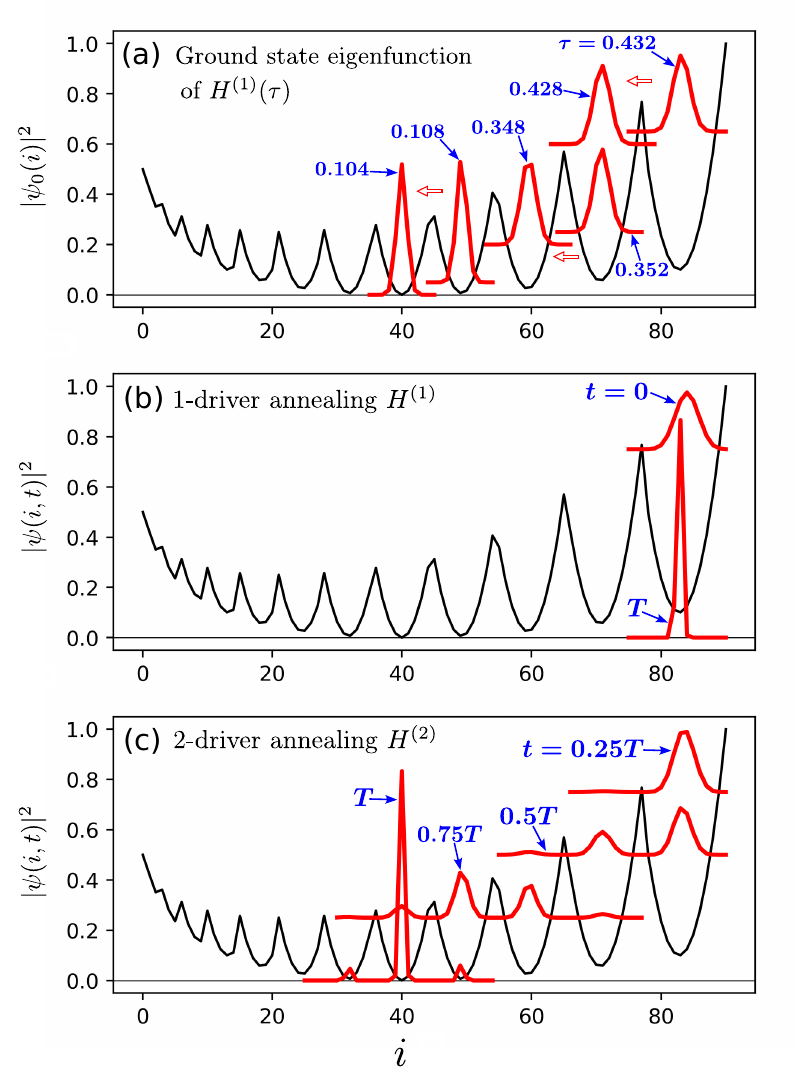}
\caption{Wave functions of Landscape I. Probability densities (red) are shifted vertically for visual clarity (heights and horizontal coordinates are faithful to the scales of both axes). Landscape I (black) is also shown. (a) Statics. Ground state of $H^{(1)}$ at $\tau$'s before and after gap closures [see Fig. \ref{fig.gaps.Landscape I}(a)], undergoing first-order transitions. (b) Time evolution of wave function, annealing under $H^{(1)}$ ($T=1000$). The system is trapped in a local minimum. (c) Annealing along 2-driver path $s(\tau)$ [see Fig. \ref{fig.gaps.Landscape I}(b)], where the system attains the global minimum (same $T$).}
\label{fig.Landscape I.psis}
\end{center}
\end{figure}

Consider the annealing Hamiltonian
\begin{equation}
H^{(1)}(\tau)
=
(1-\tau)
\,
H_{\mathrm{P}}^{\mathrm{I}}
-
\tau \, T_x
\label{eq.H(1)}
\end{equation}
where $T_x$ is the driver, and the annealing parameter $\tau\in[0,1]$ starts from 1 and ends at 0. Denote the energy gap between the ground and first excited states as
\begin{equation}
\Delta=E_1-E_0
\label{eq.Delta.energy gap. defninition}
\end{equation}
The red line (1-driver) in Fig. \ref{fig.gaps.Landscape I}(a) shows the gap of Eq. (\ref{eq.H(1)}) as a function of $\tau$. As $\tau$ decreases, the gap undergoes a series of four closures in the region between 0.5 and 0. To illustrate first-order transitional behaviors at these closures, Fig. \ref{fig.Landscape I.psis}(a) shows the probability density of the ground state eigenfunction $|\psi_0(i)|^2$ at values of $\tau$ near the zero gaps. The densities have been overlaid on the graph of Landscape I, and adjusted vertically to aid visualization. For instance, we see that as $\tau$ decreases slightly from 0.432 to 0.428, the ground state jumps from the rightmost local minimum to the adjacent one. The four zero gaps reflect the fact that the system needs to overcome four energy barriers to reach the global minimum.

Can these zero gaps be avoided if, instead of $T_x$, we drive the system with $U_x$ or $V_x$? It is straightforward to show that 
\begin{align}
e^{i\pi V_x}\,H^{(1)}(\tau)\,e^{-i\pi V_x}
&=
(1-\tau)
\,
H_{\mathrm{P}}^{\mathrm{I}}
-
\tau \, U_y
\nonumber \\
&=
e^{-i\pi V_3}
\left[
(1-\tau)
\,
H_{\mathrm{P}}^{\mathrm{I}}
-
\tau \, U_x
\right]
\,e^{i\pi V_3}
\label{eq.Landscape I.Hausdorffs}
\end{align}
where $U_y=\frac{1}{2i}(U_+-U_-)$. Combining $H_{\mathrm{P}}^{\mathrm{I}}$ with either $T_x$, $U_y$, or $U_x$ gives rise to unitarily equivalent Hamiltonians, and so have the same energy gaps. Replacing $T_x$ by $V_x$ results in an inequivalent Hamiltonian, but we found that the energy gap still exhibits a series of closures, so the problem persists.

We now subject $H^{(1)}(\tau)$ to a second driver $U_x$,
\begin{equation}
H^{(2)}(\tau,s)
=
(1-s)
\,
H^{(1)}(\tau)
-
s
\,
U_x
\label{eq.H(2)}
\end{equation}
The Hamiltonian now depends on two parameters $\tau$ and $s$, with $s\in[0,1]$. This prescription for utilizing two drivers was proposed by Seki and Nishimori in the context of using antiferromagnetic interaction as a second driver to overcome first-order transitions in QA \cite{Seki12}. They showed that gap closures can be avoided via an appropriate choice of annealing path in $s$-$\tau$ space. Figure \ref{fig.gaps.Landscape I}(b) shows the gap landscape of $H^{(2)}(\tau,s)$. It is plotted in the form of a heat map, where the color at the point $(\tau,s)$ indicates the value of the gap $\Delta(\tau,s)$. The objective is to travel from the southeastern corner (start) to the southwestern corner (end). Cool colors represent small gap values. The annealing bottlenecks are shown by the blue-colored regions in the lower part of the diagram. The annealing path of the 1-driver system Eq. (\ref{eq.H(1)}) is shown by the dashed horizontal line (yellow), which traverses the blue regions. It is possible to circumvent the bottleneck region by making a detour, say, along the dashed curve labeled $s(\tau)$ (curated manually by hand). For clarity, the gap along this 2-driver path is plotted in Fig. \ref{fig.gaps.Landscape I}(a) as the dashed curve (blue). We see that the energy gap does not close along this path. 

\subsection{Nonlocal transport during quantum annealing}
\label{subsec.Annealing.Landscape I}

We now consider annealing. The time-dependent Schr\"odinger equation is propagated numerically from $t=0$ to $T$, where $t$ denotes the time variable and $T$ is the total annealing time \cite{footnote.01}. We set $\hbar=1$ in all calculations. The initial wave function is the ground state of the annealing Hamiltonian at $(\tau,s)=(1,0)$. We move along the annealing path in such a way that $t$ is evenly spaced along the entire curve. In other words, $\frac{t}{T}$ is the fractional arclength that has been traversed along the curve at time $t$. For 1-driver annealing, this becomes 
\begin{equation}
\tau=1-\frac{t}{T}
\label{eq.linear schedule}
\end{equation}
which is the familiar linear schedule. For 2-driver annealing, the above procedure is implemented numerically.

Figure \ref{fig.Landscape I.psis}(b) shows the time evolution of the probability density $|\psi(i,t)|^2$ under the 1-driver Hamiltonian Eq. (\ref{eq.H(1)}), with $T=1000$. The wave functions at $t=0$ and $T$ are shown. The system is trapped in a local minimum, and this behavior persists for longer $T$'s. Panel (c) shows the results for 2-driver annealing under Eq. (\ref{eq.H(2)}) along the path $s(\tau)$ in Fig. \ref{fig.gaps.Landscape I}(b), with the same $T$. In this case, the wave function attains the global minimum. From the profiles of the probability density curves, it seems that tunneling is not involved in the process, because the wave function in the barrier regions is negligible. Rather, it appears as if the system is transported directly to the adjacent minima without passing through the classically-forbidden region. Such nonlocal transports could be seen as another mechanism---apart from adiabaticity and tunneling---that QA can employ during the process of optimization. We wish to emphasize that this property arises from the irreducible representation of the $T_x$ and $U_x$ operators, and is a natural feature of $\mathfrak{su}(3)$ algebra.

To assess the results of annealing, we consider the residual energy, defined as 
\begin{equation}
R(T)=\langle \psi(T) |H_{\mathrm{P}}| \psi(T) \rangle - E_{\mathrm{G}}
\label{eq.Residual Energy.definition}
\end{equation}
where $\psi(T)$ is the wave function at time $t=T$, and $E_{\mathrm{G}}$ is the ground state energy of the problem Hamiltonian $H_{\mathrm{P}}$. Equation (\ref{eq.Residual Energy.definition}) is a measure of how quickly QA is able to attain the global minimum. If $R(T)$ decays to zero quickly with respect to $T$, then QA is an efficient algorithm. If it decreases slowly or stagnates, then it is an indication that QA is ineffective or has failed. 

\begin{figure}[h]
\begin{center}
\includegraphics[scale=0.4]{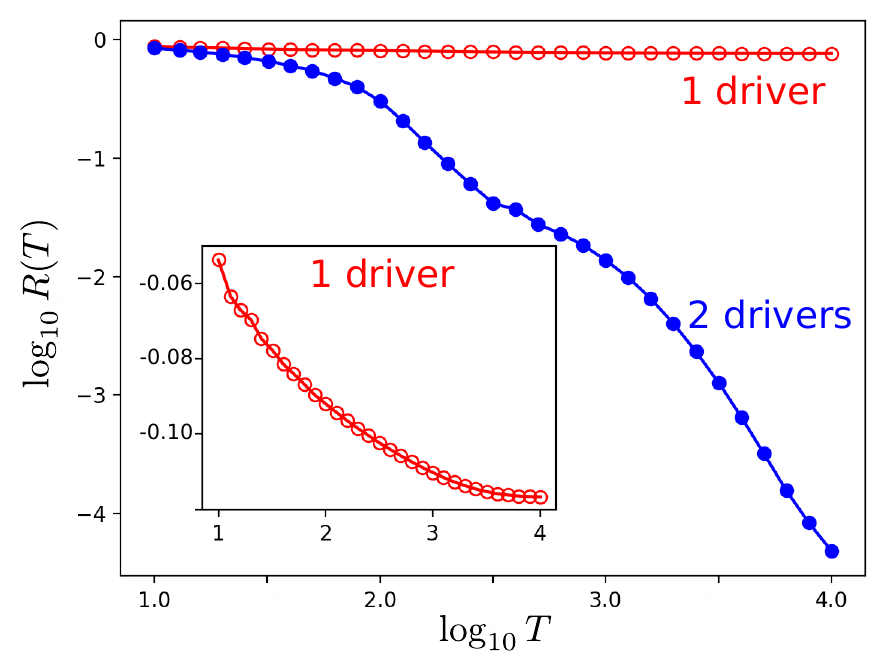}
\caption{Residual energies of Landscape I. Open circles (red) show $R(T)$ along the 1-driver path [Fig. \ref{fig.gaps.Landscape I}(b)]. Solid circles (blue) show that along the 2-driver path $s(\tau)$. The former shows almost no decrease due to trappings by local minima, while the latter shows a steady decay. Inset: The 1-driver curve in detail. Connecting lines are to guide the eye only.}
\label{fig.REvsT.Landscape I}
\end{center}
\end{figure}

Figure \ref{fig.REvsT.Landscape I} shows the results for Landscape I. Open circles (red) show the $R(T)$ curve obtained from annealing along the 1-driver path in Fig. \ref{fig.gaps.Landscape I}(b). The residual energy decreases extremely slowly, which is to be expected from our discussion of the time evolution of its wave function above. Solid circles (blue) show the results along the 2-driver path $s(\tau)$. In this case, the residual energy shows a steady decrease. We also studied how multiplet size affects the $R(T)$ curve along the 2-driver path, increasing $(p,0)$ from $p=13$ to 22. In general, the results for $p=12$ discussed above are representative of larger multiplets. Presentation of these results will be deferred to Sec. \ref{sec.Jx only} when we compare them with traditional QA.


\section{Landscape II: Annealing dynamics in multiplet space}
\label{sec.landscape II}

Landscape II is obtained by inverting Landscape I, i.e. $H_{\mathrm{P}}^{\mathrm{II}}=-H_{\mathrm{P}}^{\mathrm{I}}$. Figure \ref{fig.4 su3 potentials}(b) shows the graph of the landscape for $(p,q)=(12,0)$. It is shaped as a concave envelope, with local minima along its surface. The envelope represents a large barrier separating two deep energy basins. In the Hopfield model, when the system is trapped in local minima known as spurious states, one must transcend a high barrier to escape \cite{Nishimori96}. Such situations also arise in the study of transition states \cite{Kumeda01}. Landscape II simulates these scenarios. Note that here we are looking at inter-basin annealing, whereas Landscape I is concerned with intra-basin minimization.

\begin{figure}[h]
\begin{center}
\includegraphics[scale=0.45]{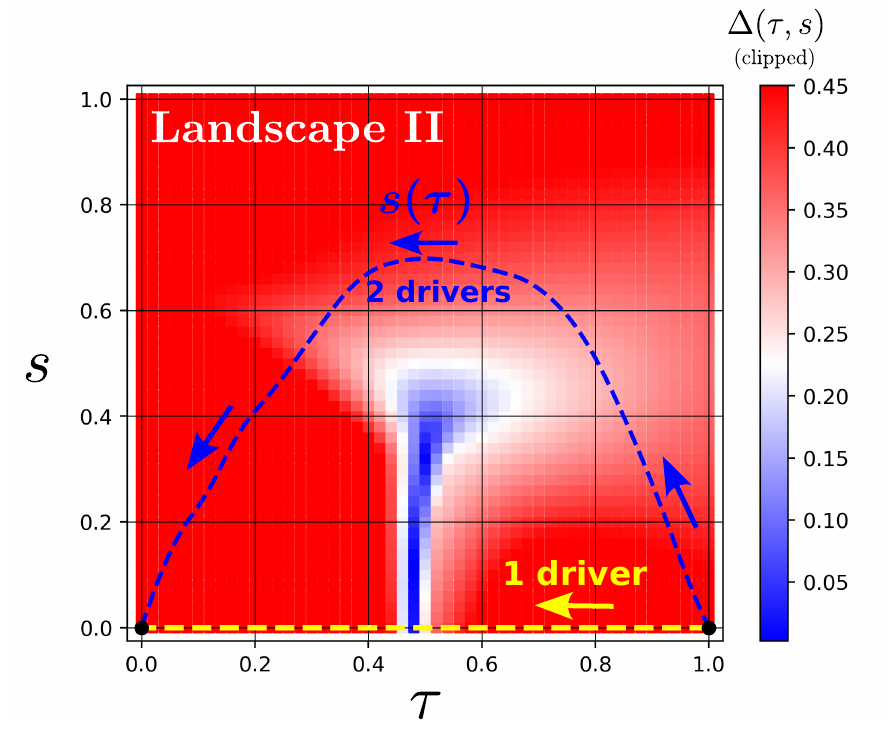}
\caption{Gap landscape of $H^{(3)}(\tau,s)$ [Eq. (\ref{eq.H(3)})], organized similar to Fig. \ref{fig.gaps.Landscape I}(b). There is a region of gap closure at $\tau\approx 0.478$ (blue vertical strip), which is lifted when $s>0.5$. The 1-driver path traverses this region, while the 2-driver path avoids it.}
\label{fig.gap landscape.Landscape II}
\end{center}
\end{figure}

Analogous to Eq. (\ref{eq.H(2)}), we define a 2-driver Hamiltonian 
\begin{equation}
H^{(3)}(\tau,s)=
(1-s)
\left[
(1-\tau)
\,
H_{\mathrm{P}}^{\mathrm{II}}
-
\tau
\,
V_x
\right]
-
s
\,
U_x
\label{eq.H(3)}
\end{equation}
where $\tau,s\in[0,1]$ as before. When $s=0$, one has a 1-driver system driven by $V_x$. Here, we do not use $T_x$ as the primary driver [unlike Eq. (\ref{eq.H(1)})] because its ground state is trivially close to the global minimum of Landscape II [compare Figs. \ref{fig.4 su3 potentials}(b) and \ref{fig.Landscape I.psis}(b)]. Figure \ref{fig.gap landscape.Landscape II} shows the gap landscape of $H^{(3)}(\tau,s)$. The salient feature is a vertical strip of gap closure (blue region). The annealing path of the 1-driver system is indicated by the horizontal path (yellow), which traverses the zero gap region at $\tau\approx 0.48$. 

\begin{figure}[h]
\begin{center}
\includegraphics[scale=0.45]{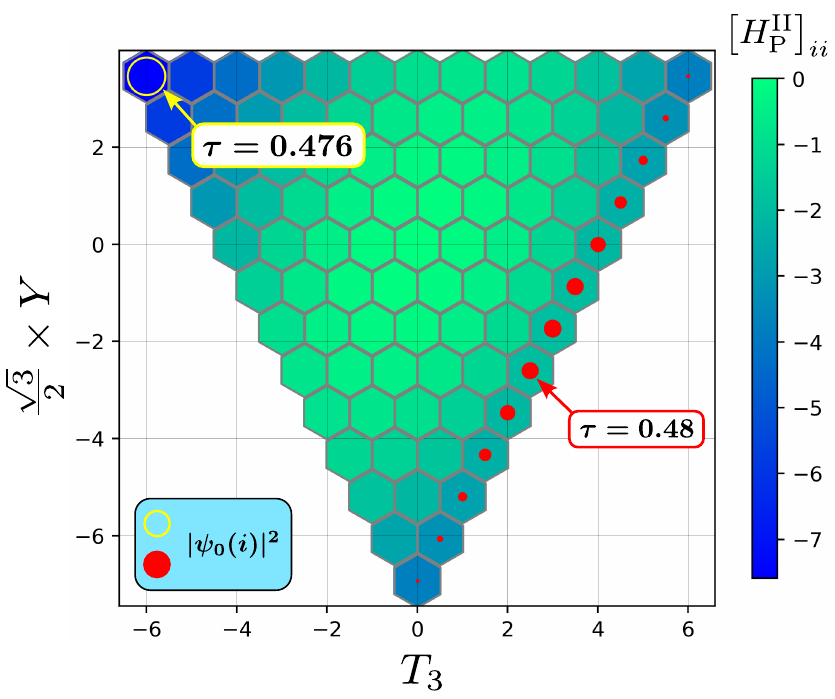}
\caption{Ground state eigenfunction of $H^{(3)}$ along the 1-driver path (Fig. \ref{fig.gap landscape.Landscape II}), overlaid on Landscape II in multiplet space. The color of each hexagonal cell represents the matrix element $[H_{\mathrm{P}}^{\mathrm{II}}]_{ii}$, while circle symbols indicate the probability density $|\psi_0(i)|^2$. The index $i$ is shown in Fig. \ref{fig.Landscape II.su(3) space} and explained in Appendix \ref{app.sec.irrep of su(3)}. The ground state jumps abruptly over the energy barrier (greenish region), characteristic of a first-order transition.}
\label{fig.Landscape II.gds density.statics}
\end{center}
\end{figure}

Hitherto, we have been viewing the energy landscapes in configuration space. The $\mathfrak{su}(3)$ framework offers another perspective in multiplet space. Figure \ref{fig.Landscape II.gds density.statics} shows Landscape II plotted as a two-dimensional heat map on the $(12,0)$-multiplet diagram [Fig. \ref{fig.su3 YvsT3 plot}(a)]. The color of a hexagonal cell indicates the value of the matrix element $[H_{\mathrm{P}}^{\mathrm{II}}]_{ii}$, where the mapping between the index $i$ and coordinates $(T_3,Y)$ is discussed in Appendix \ref{app.sec.irrep of su(3)}.

To understand the annealing bottleneck in Fig. \ref{fig.gap landscape.Landscape II}, let us examine the ground state eigenfunction in multiplet space. Figure \ref{fig.Landscape II.gds density.statics} shows its probability density $|\psi_0(i)|^2$ along the 1-driver path, overlaid on Landscape II. The $i$th element is plotted in the corresponding hexagon as a circle, with area proportional to the value of $|\psi_0(i)|^2$. The eigenfunctions at $\tau=0.48$ (solid circles) and 0.476 (open circle) are shown, corresponding to the ground state just before and after the gap closure along the 1-driver path. We see that the ground state jumps abruptly across the energy barrier (greenish region), characteristic of a first-order transition. When annealing along the 1-driver path, we found that the wave function is trapped in hexagons $i=0$ and 78 (see Fig. \ref{fig.Landscape II.su(3) space}).

\begin{figure*}
\begin{center}
\includegraphics[scale=0.35]{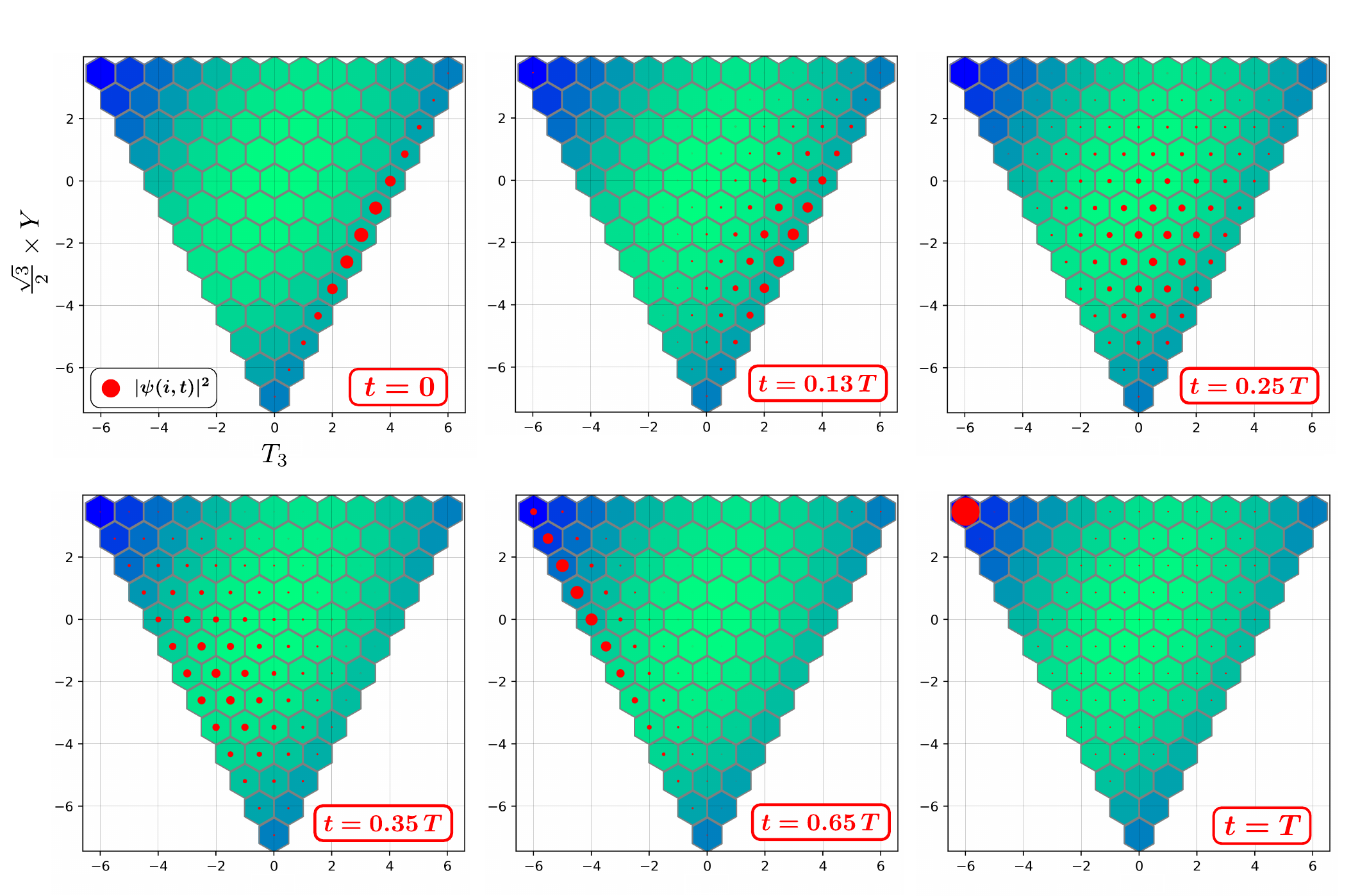}
\caption{Time evolution of probability density $|\psi(i,t)|^2$ when annealing along the 2-driver path in Fig. \ref{fig.gap landscape.Landscape II}. Each panel shows the snapshot at time $t$ ($T=1000$), and is organized similar to Fig. \ref{fig.Landscape II.gds density.statics}. The wave function crosses the energy barrier and attains the global minimum.}
\label{fig.Landscape II.psi(i,t).two rows}
\end{center}
\end{figure*}

The gap closure can be avoided by following the 2-driver path $s(\tau)$ in Fig. \ref{fig.gap landscape.Landscape II} (blue, curated manually). Figure \ref{fig.Landscape II.psi(i,t).two rows} shows the time evolution of probability density $|\psi(i,t)|^2$ in multiplet space when annealing along this path. Each panel is organized similar to Fig. \ref{fig.Landscape II.gds density.statics}, and shows the snapshot at time $t$ ($T=1000$). We see that the wave function crosses the energy barrier and attains the global minimum. The underlying mechanism can be understood based on the actions of the quantum drivers $V_x$ and $U_x$. From Fig. \ref{fig.su3 YvsT3 plot}(a), we know that the operators $V_{\pm}$ translate the wave function along $60^{\circ}$ polar angle in multiplet space, while $U_{\pm}$ act along the $120^{\circ}$ angle. Their combined actions enable the wave function to explore the entire multiplet diagram thoroughly, greatly improving the chances of finding the global minimum. Note that this intuitive explanation would not be apparent if we had viewed things in configuration space [i.e. Fig. \ref{fig.4 su3 potentials}(b)].

Similar to Landscape I, we also calculated the residual energy along the 2-driver path, and investigated its dependence on multiplet size. It was found that $R(T)$ decreases asymptotically as $\sim T^{-2}$ for $(p,0)$ from $p=12$ to 22. This adiabatic scaling \cite{Morita07, Suzuki05} is indication that the annealing is effective. We shall defer presentation of these results to Sec. \ref{sec.Jx and another} when we make comparisons with other annealing models.


\section{Landscape III: Multi-layered energy surface}
\label{sec.landscape III.pure}

\begin{figure}[h]
\begin{center}
\includegraphics[scale=0.45]{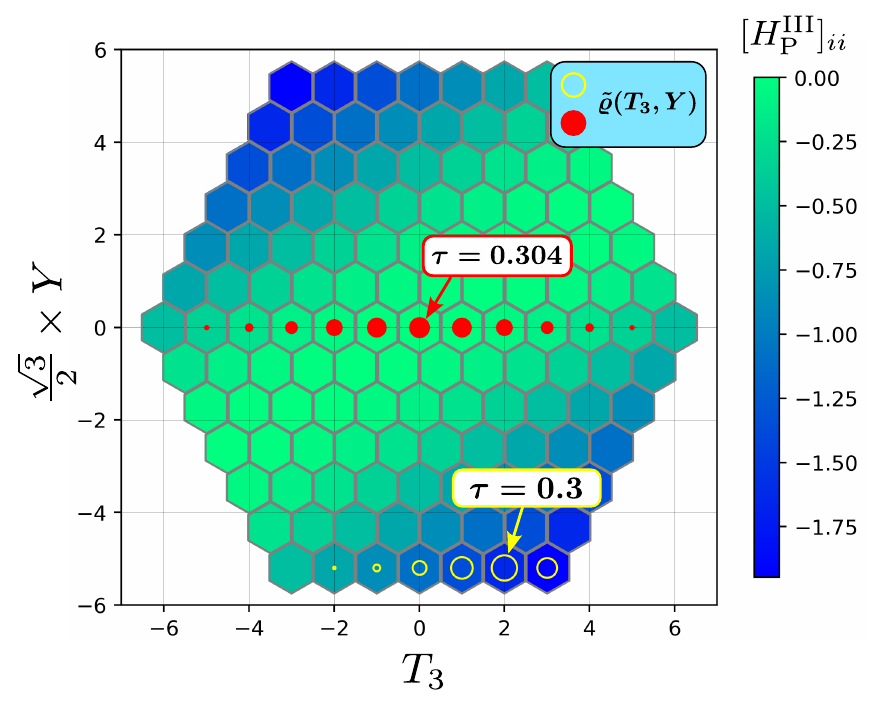}
\caption{Ground state eigenfunction of $H^{(4)}$ along the 1-driver path [Fig. \ref{fig.gap landscape.Landscape III.only su3 drivers}(a)], overlaid on Landscape III (Layer 7) in multiplet space. Circles indicate the cumulated probability density $\tilde{\varrho}(T_3,Y)$ [Eq. (\ref{eq.P(T_3,Y).definition})]. The densities at $\tau=0.304$ (solid circles) and 0.3 (open circles) are shown. One sees that the wave function makes a first-order transition.}
\label{fig.landscapeIII.gds density.multiplet space}
\end{center}
\end{figure}

We now consider Landscape III. The problem Hamiltonian is 
\begin{equation}
H_{\mathrm{P}}^{\mathrm{III}}=-\frac{1}{\tilde{n}}U_3^2
\label{}
\end{equation}
Figure \ref{fig.4 su3 potentials}(c) shows the graph for $(p,q)=(6,6)$. This landscape can be seen as a combination of Landscapes I and II. There are deep as well as shallow minima, and so is more challenging to anneal because one must perform both intra- and inter-basin optimizations. Landscape III also brings out another aspect of the $\mathfrak{su}(3)$ framework, which is that when both $p$ and $q$ are non-zero the energy surface in multiplet space consists of more than one layer. This is due to multiplicity, discussed in Sec. \ref{sec.su2 and3 review} [see Fig. \ref{fig.su3 YvsT3 plot}(b)]. On the multiplet diagram, the problem Hamiltonian depends only on the coordinates $(T_3,Y)$, so the energy surfaces are the same on every layer. Landscape III has seven layers, and Layer 7 is plotted in Figure \ref{fig.landscapeIII.gds density.multiplet space} (like Landscape II in Fig. \ref{fig.Landscape II.gds density.statics}). This layer is representative of the entire landscape in multiplet space. (See Appendix \ref{app.sec.irrep of su(3)} for explanation of the different layers.)

\begin{figure*}
\begin{center}
\includegraphics[scale=0.45]{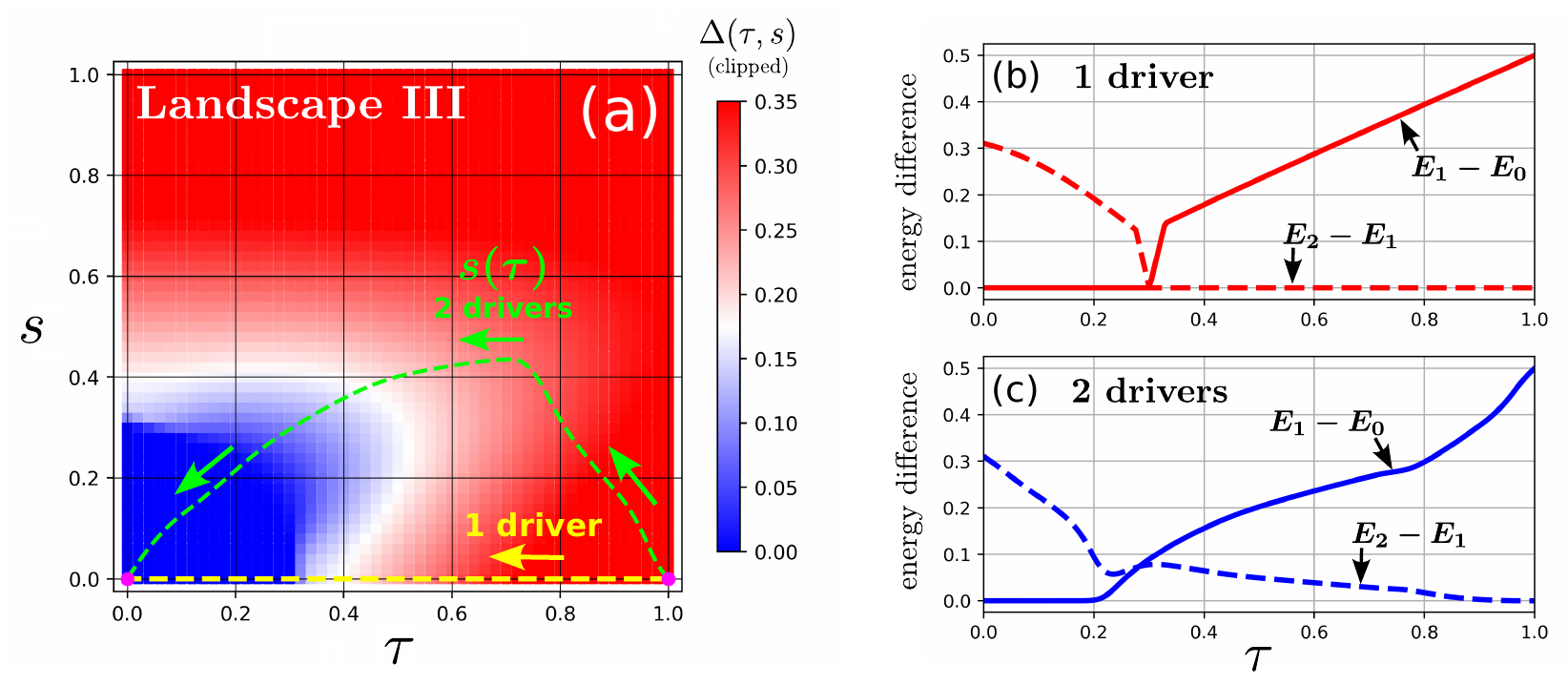}
\caption{Energy gaps of Landscape III. (a) Gap landscape of $H^{(4)}(\tau,s)$ [Eq. (\ref{eq.H(4).landscapeIII.pure})]. Along the 1-driver path, the gap closes sharply at $\tau\approx 0.3$. This is avoided along the 2-driver path, which descends along a gentler gradient (whitish region). (b) Energy gaps $E_1-E_0$ and $E_2-E_1$ along the 1-driver path. Gap closure is indicated by the `V' at $\tau\approx 0.3$. (c) The gaps along the 2-driver path $s(\tau)$. The `V' is lifted, and there is no first-order transition along this path.}
\label{fig.gap landscape.Landscape III.only su3 drivers}
\end{center}
\end{figure*}

Proceeding to annealing, define the 2-driver Hamiltonian
\begin{equation}
H^{(4)}(\tau,s)
=
(1-s)
\,
\left[
(1-\tau)
\,
H^{\mathrm{III}}_{\mathrm{P}}
-
\tau
\,
T_x
\right]
-
s
\,
V_x
\label{eq.H(4).landscapeIII.pure}
\end{equation}
where $\tau,s \in [0,1]$. When $s=0$, we have a 1-driver system driven by $T_x$. Figure \ref{fig.gap landscape.Landscape III.only su3 drivers}(a) shows the gap landscape of $H^{(4)}(\tau,s)$. The southwestern region (blue) exhibits small gap, because the ground state of $H^{\mathrm{III}}_{\mathrm{P}}$ is two-fold degenerate [see Fig. \ref{fig.4 su3 potentials}(c)]. The annealing path of the 1-driver system is indicated by the horizontal line (yellow), along which the gap closes sharply at $\tau\approx 0.3$. For clarity, the solid line in panel (b) shows the gap $E_1-E_0$ along this path. The dashed line shows the subsequent gap, $E_2-E_1$, along the same path. The salient feature is the `V'-shaped closure at $\tau\approx 0.3$ where all three levels $E_0, E_1$, and $E_2$ become degenerate. This V-closure is a genuine annealing bottleneck, not a manifestation of ground state degeneracy.
 
Figure \ref{fig.landscapeIII.gds density.multiplet space} shows the ground state eigenfunction $\psi_0(i)$ near the V-closure, overlaid on Landscape III in multiplet space. Due to the layered structure of the multiplet diagram, here the area of a circle symbol at coordinates $(T_3,Y)$ is proportional to the cumulated probability density
\begin{equation}
\tilde{\varrho}(T_3,Y)=\tilde{\sum_{i}}|\psi_0(i)|^2
\label{eq.P(T_3,Y).definition}
\end{equation}
where the sum runs over all states $i$ with eigenvalues $T_3$ and $Y$ (see Appendix \ref{app.sec.irrep of su(3)}). The physical meaning of Eq. (\ref{eq.P(T_3,Y).definition}) is that we project $\psi_0(i)$ of all seven layers onto Layer 7. This provides a compact view of the entire wave function, but at the expense of losing information about individual layers. The figure shows $\tilde{\varrho}$ at $\tau=0.304$ and 0.3, which are slightly before and after the V-closure. It is seen that the wave function makes a jump in multiplet space---a first-order transition. When annealing along the 1-driver path, we found that the wave function is trapped in hexagons $i=330$ and 342 [see Fig. \ref{fig.Landscape III.su(3) multiplet space}(b)].

\begin{figure*}
\begin{center}
\includegraphics[scale=0.35]{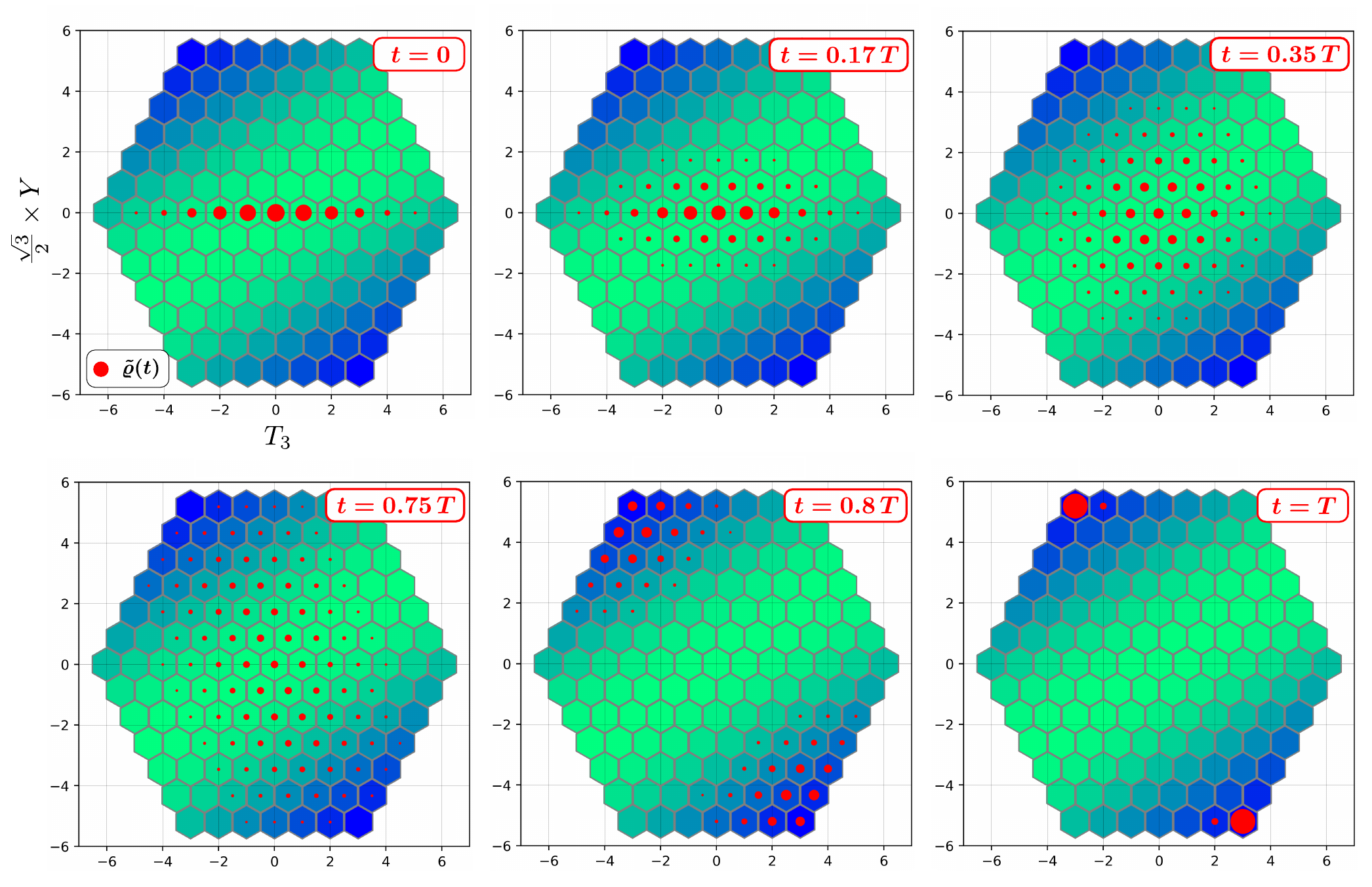}
\caption{Time evolution of cumulated density $\tilde{\varrho}(t)$ when annealing along the 2-driver path in Fig. \ref{fig.gap landscape.Landscape III.only su3 drivers}(a). Each panel shows the snapshot at time $t$ ($T=1000$), and is organized similar to Fig. \ref{fig.landscapeIII.gds density.multiplet space}. The wave function bifurcates continuously and attains the two global minima.}
\label{fig.landscapeIII.dynamics.stacked.1and2driver}
\end{center}
\end{figure*}

To address the V-closure, one needs to find a suitable path on the gap landscape. For Landscapes I and II, this is relatively straightforward because we just need to avoid the blue regions [see Figs. \ref{fig.gaps.Landscape I}(b) and \ref{fig.gap landscape.Landscape II}]. In Fig. \ref{fig.gap landscape.Landscape III.only su3 drivers}(a), however, it seems that one cannot avoid traversing the blue region to reach the southwestern corner. Nevertheless, note that there is a whitish region where the gap decreases gradually, rather than sharply. Based on this observation, we propose a 2-driver path $s(\tau)$ (green, curated manually) which descends along this gentler gradient. Figure \ref{fig.gap landscape.Landscape III.only su3 drivers}(c) shows the energy gaps $E_1-E_0$ (solid) and $E_2-E_1$ (dashed) along this path. The `V' is lifted, and the two gaps no longer close at the same point. (The gap $E_1-E_0$ still becomes zero towards the end of the path, but this is because of ground state degeneracy and does not pose problems for minimization.) 

\begin{figure*}
\begin{center}
\includegraphics[scale=0.35]{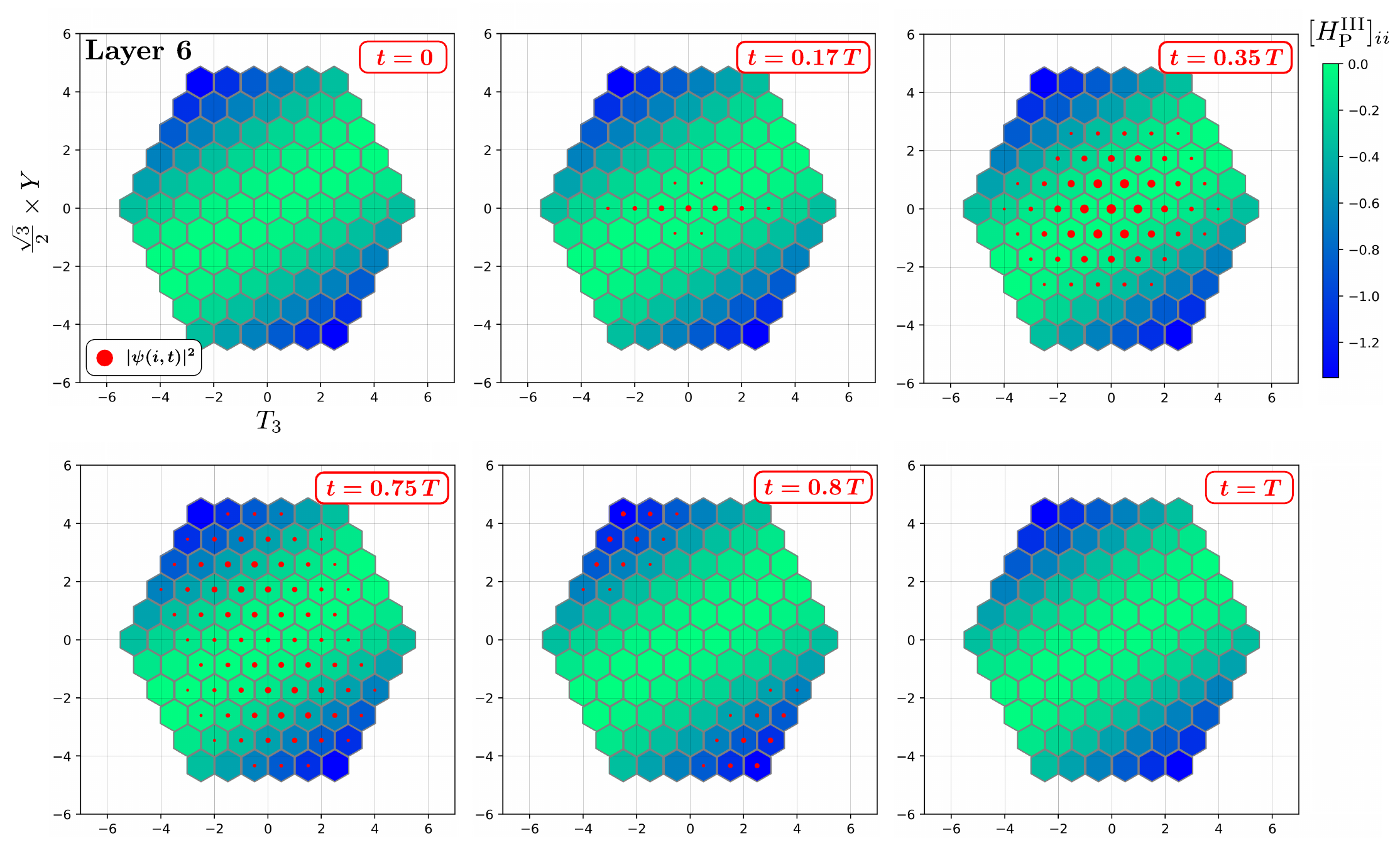}
\caption{Time evolution of probability density on Layer 6 of Landscape III, for the annealing shown in Fig. \ref{fig.landscapeIII.dynamics.stacked.1and2driver}. Here, the area of a circle symbol is proportional to $|\psi(i,t)|^2$ where $i$ lies in Layer 6. (See Appendix \ref{app.sec.irrep of su(3)} for definition of layers.) We see that the wave function diffuses in and out of the layer during the course of annealing.}
\label{fig.LandscapeIII.dynamics.Layer6only}
\end{center}
\end{figure*}

Figure \ref{fig.landscapeIII.dynamics.stacked.1and2driver} shows the time evolution of cumulated density $\tilde{\varrho}(t)$ when annealing along this path. Each panel shows the snapshot at time $t$ ($T=1000$), with the time-dependent wave function $\psi(i,t)$ in place of $\psi_0(i)$ in Eq. (\ref{eq.P(T_3,Y).definition}). We see that the wave function undergoes a continuous bifurcation and attains the two global minima. The bifurcation process is reminiscent of the ferromagnetic Ising model undergoing a second-order phase transition \cite{Botet83}. An important difference is that here the energy surface is multi-layered. Different layers are are interconnected via the ladder operators $V_{\pm}$. During annealing, the wave function can diffuse into the different layers when searching for the global minimum. To illustrate, the wave function in Fig. \ref{fig.landscapeIII.dynamics.stacked.1and2driver} is replotted in Fig. \ref{fig.LandscapeIII.dynamics.Layer6only}, this time showing the density $|\psi(i,t)|^2$ only for $i$ lying in Layer 6 of the energy surface. At the beginning $t=0$, there is no probability amplitude in Layer 6. The wave function then diffuses into this layer from the center, spreads to the low energy regions, and departs from the layer again. We may look upon the layer as an auxiliary space that facilitates the diffusion of the wave function during annealing, helping it avoid being trapped by local minima.

\begin{figure}[h]
\begin{center}
\includegraphics[scale=0.4]{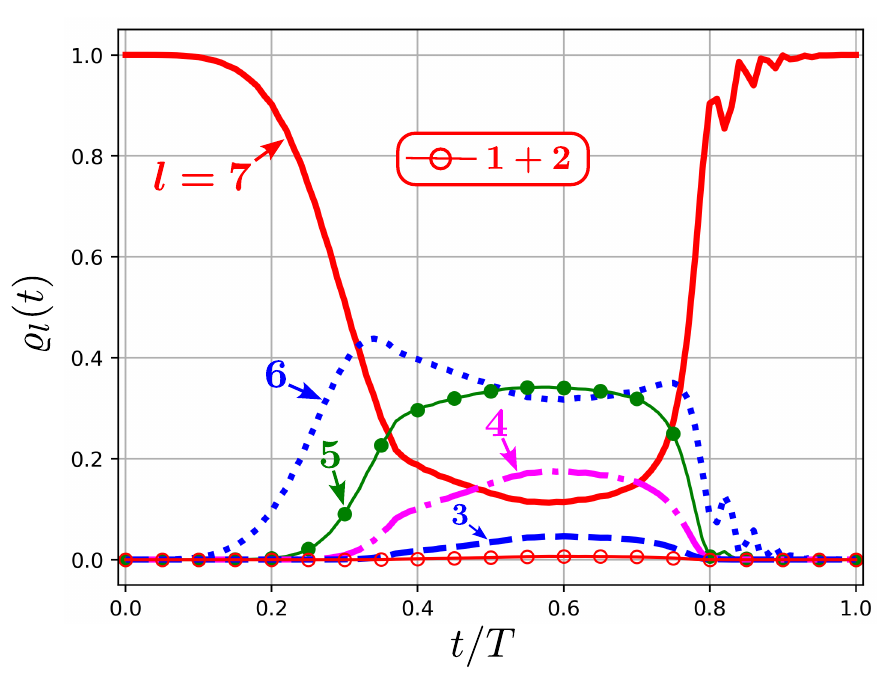}
\caption{Time evolution of layer probabilities $\varrho_l(t)$ [Eq. (\ref{eq.LandscapeIII.Probl(t)})] for the annealing shown in Fig. \ref{fig.landscapeIII.dynamics.stacked.1and2driver}. Subscript $l$ indicates layer number. A significant part of the wave function diffuses into Layers 3 to 6, particularly during the middle part of the trajectory.}
\label{fig.LandscapeIII.density in layers 1 to 7}
\end{center}
\end{figure}

To give a fuller view of how the wave function moves through the different layers, we define the probability that the system is in Layer $l$ at time $t$,
\begin{equation}
\varrho_l(t)=
\sum_{i\in \, \mathrm{Layer}\, l} |\psi(i,t)|^2
\label{eq.LandscapeIII.Probl(t)}
\end{equation}
where the sum runs through all states $i$ which are in Layer $l$. In other words, $\varrho_l(t)$ measures how much of the wave function is in Layer $l$. Figure \ref{fig.LandscapeIII.density in layers 1 to 7} shows the $\varrho_l(t)$ of all the layers, for the annealing of Fig. \ref{fig.landscapeIII.dynamics.stacked.1and2driver}. Initially, the wave function is in Layer 7. As time progresses, a significant portion diffuses into Layers 3 to 6, particularly during the middle part of annealing. Towards the end, the wave function retreats from these layers, returning back to Layer 7. Hence, we see that the structure of $\mathfrak{su}(3)$ multiplet is advantageous to QA as there are different channels for the wave function to explore the energy landscape.

We also studied the dependence of energy gaps and residual energy on multiplet size, for $(p,p)$-multiplet from $p=4$ to 11. The lifting of the V-closure continues to hold for higher multiplets. The $R(T)$ curves were found to decay diabatically (i.e., not $\sim T^{-2}$), and there was no stagnation in the residual energy. Overall, the success of 2-driver annealing reported above is representative in general. The results for $R(T)$ will be presented later in Sec. \ref{sec.Jx and another}.


\section{Annealing the Landscapes using $J_x$ as driver}
\label{sec.Jx only}

Traditionally, QA is performed using the angular momentum operator $J_x$ as driver, which represents a transverse field. It might be helpful to study how the $\mathfrak{su}(3)$ framework performs in comparison with conventional QA. Let us define the following annealing Hamiltonian for Landscape I with $J_x$ serving as driver
\begin{equation}
H^{(5)}(\tau)
=
(1-\tau)
\,
H_{\mathrm{P}}^{\mathrm{I}}
-
\tau
\left(
\frac{J_x}{\tilde{n}}
\right)
\label{eq.H(5).Jx only}
\end{equation}
where $\tau\in[0,1]$. Similarly for Landscapes II and III, define $H^{(6)}(\tau)$ and $H^{(7)}(\tau)$ according to Eq. (\ref{eq.H(5).Jx only}), with $H_{\mathrm{P}}^{\mathrm{II}}$ and $H_{\mathrm{P}}^{\mathrm{III}}$ in place of $H_{\mathrm{P}}^{\mathrm{I}}$, respectively. The operator $J_x$ is in the standard $\mathfrak{su}(2)$ irreducible representation, with $\hbar=1$ and spin $j=\frac{d_m-1}{2}$ where $d_m$ is the dimension of the problem Hamiltonian it is driving. The factor $\tilde{n}$ is again to ensure a proper scaling of energies. In using $J_x$ as driver, one is no longer working in $\mathfrak{su}(3)$ multiplet space because $J_x$ is not an element of $\mathfrak{su}(3)$. Hence, we lose the physical interpretation in terms of a system of interacting qutrits (or Kerr oscillators) in which the Casimir invariant is conserved. Nevertheless, Eq. (\ref{eq.H(5).Jx only}) can still be understood as a one-dimensional tight-binding lattice model
\begin{align}
H^{(5)}(\tau)
&
=
(1-\tau)
\left[
\sum_i
[H^{\mathrm{I}}_{\mathrm{P}}]_{ii}
\,
|i\rangle \langle i|
\right]
\nonumber\\
&-
\frac{\tau}{\tilde{n}}
\left[
\sum_i
[J_x]_{i,i+1}\,
(
|i\rangle \langle i+1|
+
|i+1\rangle \langle i|
)
\right]
\label{eq.H(5).tight binding}
\end{align}
where the index $i$ now denotes a site, $|i\rangle$ denotes a single-particle state at site $i$, $[H^{\mathrm{I}}_{\mathrm{P}}]_{ii}$ is the energy of $|i\rangle$, and $[J_x]_{i,i+1}$ induces hopping between nearest neighbor sites [see Fig. \ref{fig.Jx and Jx2.heatmap}]. A similar model has previously been studied by Suzuki and Okada in the context of residual energy after slow annealing \cite{Suzuki05}. To be consistent with our earlier discussion, we shall still use $\mathfrak{su}(3)$ terminologies when referring to the models Eq. (\ref{eq.H(5).Jx only}).

\begin{figure*}
\begin{center}
\includegraphics[scale=0.5]{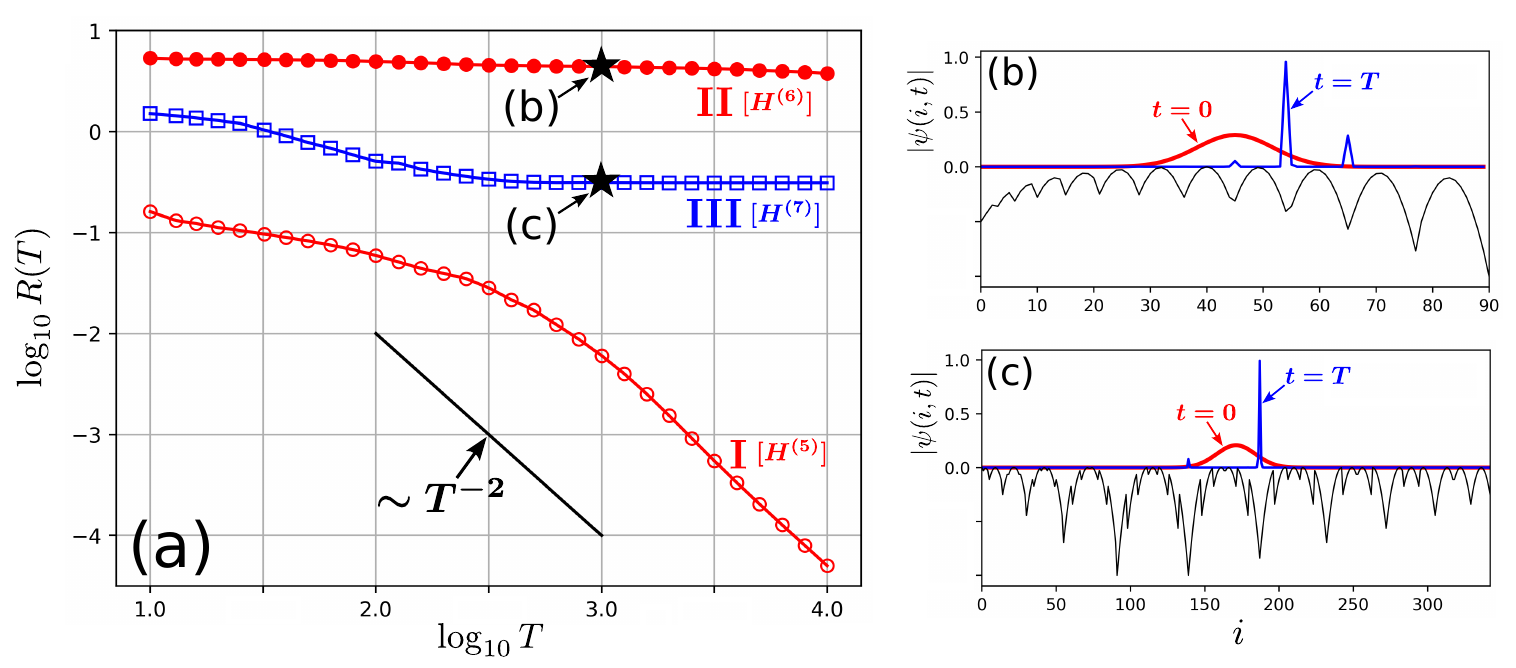}
\caption{(a) Residual energies for annealings of Landscapes I to III using $J_x$ as driver. The $R(T)$ of I (open circles) shows a steady decrease, while those of II and III stagnate. Connecting lines are to guide the eye only. Panels (b) and (c): Initial (red) and final (blue) wave functions for $T=10^3$ on the $R(T)$ curves of II and III [stars in (a)], showing the systems trapped in local minima.}
\label{fig.Jx only.RT I to II.with psi finals}
\end{center}
\end{figure*}

\begin{figure}[h]
\begin{center}
\includegraphics[scale=0.45]{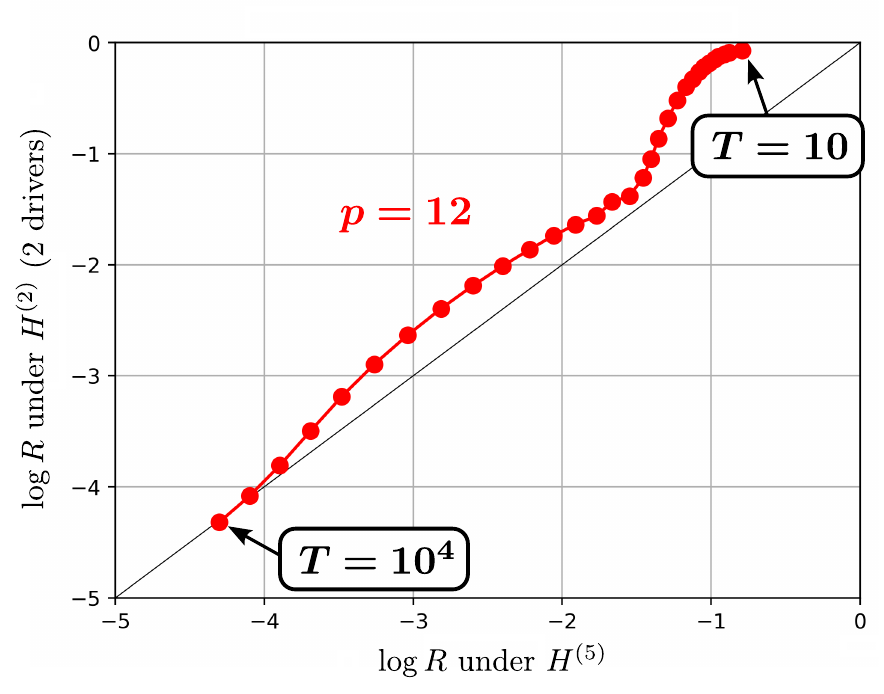}
\caption{Comparing the residual energies for Landscape I, annealing under $H^{(5)}$ (horizontal axis) versus under $H^{(2)}$ (vertical). Each point plots the results for a particular $T$, with $T$ increasing down the curve. When $T$ is large the two Hamiltonians yield similar performances. Lines connecting data points are to guide the eye only.}
\label{fig.REvsRE.LandscapeI.p=12}
\end{center}
\end{figure}

The residual energy for Landscape I annealing under $H^{(5)}$ is shown in Fig. \ref{fig.Jx only.RT I to II.with psi finals}(a) by the curve with open circles. It decreases and approaches the adiabatic limit $\sim T^{-2}$ \cite{Morita07, Suzuki05}, a sign that the annealing is successful in finding the global minimum. Examination of the energy gap reveals that there are no closures. It is instructive to compare this curve with the one for 2-driver $H^{(2)}$ in Fig. \ref{fig.REvsT.Landscape I} to see which performs better, $J_x$ or $\mathfrak{su}(3)$ drivers. Figure \ref{fig.REvsRE.LandscapeI.p=12} compares the two residual energies, where each point represents the $R(T)$ obtained under $H^{(2)}$ (vertical axis) versus that obtained under $H^{(5)}$ (horizontal), with $T$ being the same. The data points of $T=10$ and $10^4$ are indicated by arrows, with $T$ increasing (exponentially) down the curve. The diagonal line (black) denotes equality between the two $R(T)$'s. It is seen that the two yield similar performances in the sense that the curve approaches the equality line when $T$ is large.

We also see that $H^{(5)}$ gives a smaller residual energy than $H^{(2)}$ when $T$ is small. This is because the initial wave function of $H^{(5)}$ is closer to the global minimum of Landscape I. The curve labeled $t=0$ (red) in Fig. \ref{fig.Jx only.RT I to II.with psi finals}(b) shows the ground state of $J_x$, which if overlaid upon the graph of Landscape I in Fig. \ref{fig.4 su3 potentials}(a), straddles the global minimum at $i=40$. By contrast, the initial wave function of $H^{(2)}$ is quite far away, which we have already seen in Fig. \ref{fig.Landscape I.psis}(b). Hence, the early win by $H^{(5)}$ is not necessarily a sign that $\mathfrak{su}(3)$ is inferior, but simply a consequence of initial conditions. One should focus on the large $T$ behavior to make an assessment.

\begin{figure*}
\begin{center}
\includegraphics[scale=0.45]{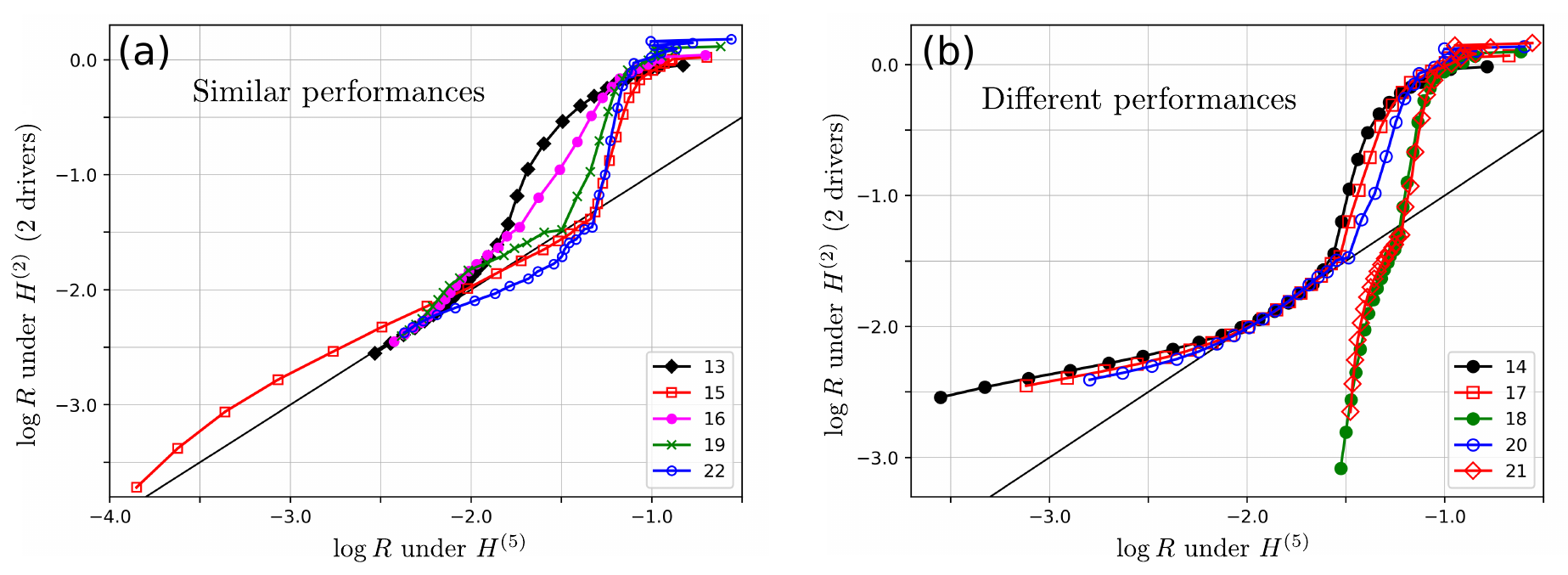}
\caption{Effects of multiplet size on the $R(T)$ vs $R(T)$ curves of Landscape I. Legends indicate the $p$ value of the multiplet. All curves start at $T=10$ and end at $T=10^4$ (as in Fig. \ref{fig.REvsRE.LandscapeI.p=12}). Connecting lines are to guide the eye only. (a) Cases where $H^{(2)}$ and $H^{(5)}$ have similar performances. (b) Cases where the performances are different.}
\label{fig.REvsRE.LandscapeI.p=13 to 22}
\end{center}
\end{figure*}

Figure \ref{fig.REvsRE.LandscapeI.p=12} compared $H^{(5)}$ with $H^{(2)}$ for the (12,0)-multiplet. We now examine how increasing the multiplet size affects these results. Figure \ref{fig.REvsRE.LandscapeI.p=13 to 22} shows the $R(T)$ versus $R(T)$ plots for Landscape I, this time for $(p,0)$-multiplets from $p=13$ to 22. The $p$ value corresponding to each curve is indicated in the legends. The results are separated into two groups. Panel (a) shows the cases where the two Hamiltonians exhibit similar performances, like in Fig. \ref{fig.REvsRE.LandscapeI.p=12}. On the other hand, we also encountered cases where they give different residual energies for large $T$, and these are shown in (b). For $p=18$ and 21, $\mathfrak{su}(3)$ drivers perform better than $J_x$; for $p=14, 17$, and 20, the reverse is true. We also examined the energy gap of $H^{(5)}$ and found it to be inconsistent in the sense that for some $p$'s it exhibits closure while for others it does not. This is in contrast to that of $H^{(2)}$, which remain gapped with increase in multiplet size. Overall, our results indicate that annealing Landscape I using $J_x$ as driver is somewhat unpredictable in terms of efficacy, and its performance relative to $\mathfrak{su}(3)$ drivers depends erratically on multiplet size.
 
For Landscapes II and III, the results are in favor of $\mathfrak{su}(3)$ drivers. Returning to Fig. \ref{fig.Jx only.RT I to II.with psi finals}(a), we see that the $R(T)$ curves for $H^{(6)}$ and $H^{(7)}$ stagnate. For both Hamiltonians, the final annealed wave functions for $T=10^3$, indicated by stars on the $R(T)$ curves, are shown by the blue curves ($t=T$) in panels (b) and (c). It is seen that they are trapped in local minima. Note that the initial wave functions (red) are now displaced from the global minima, unlike in Landscape I. More generally, examination of the energy gaps of $H^{(6)}$ (for $12 \le p\le 22$) and $H^{(7)}$ (for $4 \le p\le 11$) revealed gap closures for almost every multiplet. In contrast, those of $H^{(3)}$ and $H^{(4)}$ along 2-driver paths remain gapped with increasing multiplet size. Hence, $\mathfrak{su}(3)$ drivers will perform better than $J_x$ in minimizing Landscapes II and III. 


\section{Annealing with $J_x$ and a SU(3) driver}
\label{sec.Jx and another}

\begin{figure*}
\begin{center}
\includegraphics[scale=0.5]{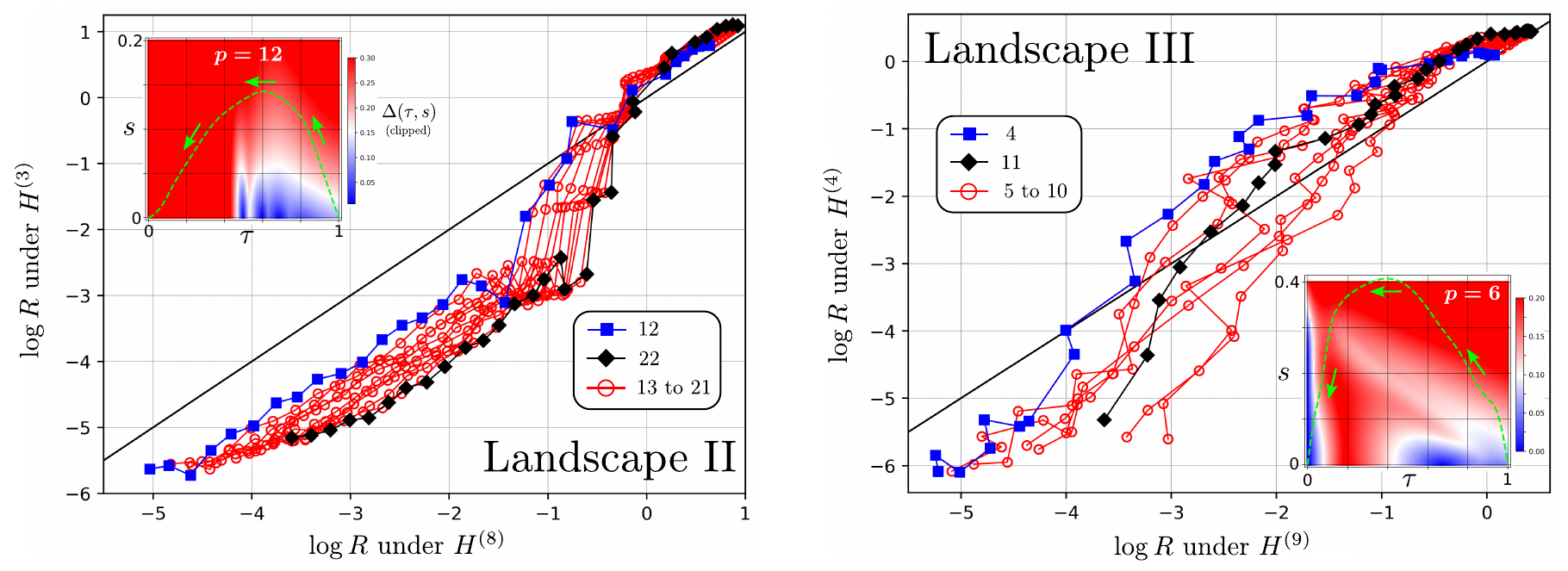}
\caption{Comparing the performances of different annealing Hamiltonians for Landscapes II (left) and III (right). Each panel shows the $R(T)$ vs $R(T)$ curves in the manner of Fig. \ref{fig.REvsRE.LandscapeI.p=13 to 22}. Legends indicate the $p$ values of the curves. For both Landscapes, two $\mathfrak{su}(3)$ drivers perform better than $J_x$ and $U_x$. Lines connecting data points are to guide the eye only. Insets show the gap landscapes of $H^{(8)}$ and $H^{(9)}$, with dashed curves (green) indicating the annealing paths. The $R(T)$'s of $H^{(8)}$ and $H^{(9)}$ for different $p$'s are obtained along these two paths.}
\label{fig.REvsRE.IIandIII.comparison}
\end{center}
\end{figure*}

In view of the difficulties experienced by $J_x$ in annealing Landscapes II and III, let us try to improve it by adding a $\mathfrak{su}(3)$ driver. Consider supplementing $H^{(6)}$ (Landscape II) with $U_x$
\begin{equation}
H^{(8)}(\tau,s)=(1-s) \, H^{(6)}(\tau) - s\, U_x
\label{eq.H(8)}
\end{equation}
where $\tau,s \in [0,1]$ as before. Similarly for Landscape III, define $H^{(9)}(\tau,s)$ according to Eq. (\ref{eq.H(8)}) with $H^{(7)}$ in place of $H^{(6)}$. The insets in Fig. \ref{fig.REvsRE.IIandIII.comparison} show the gap landscapes of $H^{(8)}$ with $(p,q)=(12,0)$ and $H^{(9)}$ with $(p,q)=(6,6)$. The dashed curves (green) show annealing paths which avoid small gap regions. Along the two paths, the energy gap is gapped for $H^{(8)}$ with $12 \le p\le 22$ and $H^{(9)}$ with $4 \le p\le 11$. In other words, the performance of $J_x$ is significantly enhanced by $U_x$. Here, we would like to focus on the residual energies and see how they fare in comparison with our earlier $\mathfrak{su}(3)$ models. In Fig. \ref{fig.REvsRE.IIandIII.comparison}, the left panel compares $H^{(3)}$ with $H^{(8)}$ for Landscape II and the right panel compares $H^{(4)}$ with $H^{(9)}$ for Landscape III. All annealings are along 2-driver paths. As in Fig. \ref{fig.REvsRE.LandscapeI.p=13 to 22}, each  line represents the $R(T)$ versus $R(T)$ curve of one multiplet, whose $p$ value is indicated in the legends. To avoid cluttering the figure, many of the curves are labeled using the same symbols. For both landscapes, annealing with two $\mathfrak{su}(3)$ drivers is better than annealing with $J_x$ and $U_x$, resulting in smaller residual energies.

\begin{figure*}
\begin{center}
\includegraphics[scale=0.45]{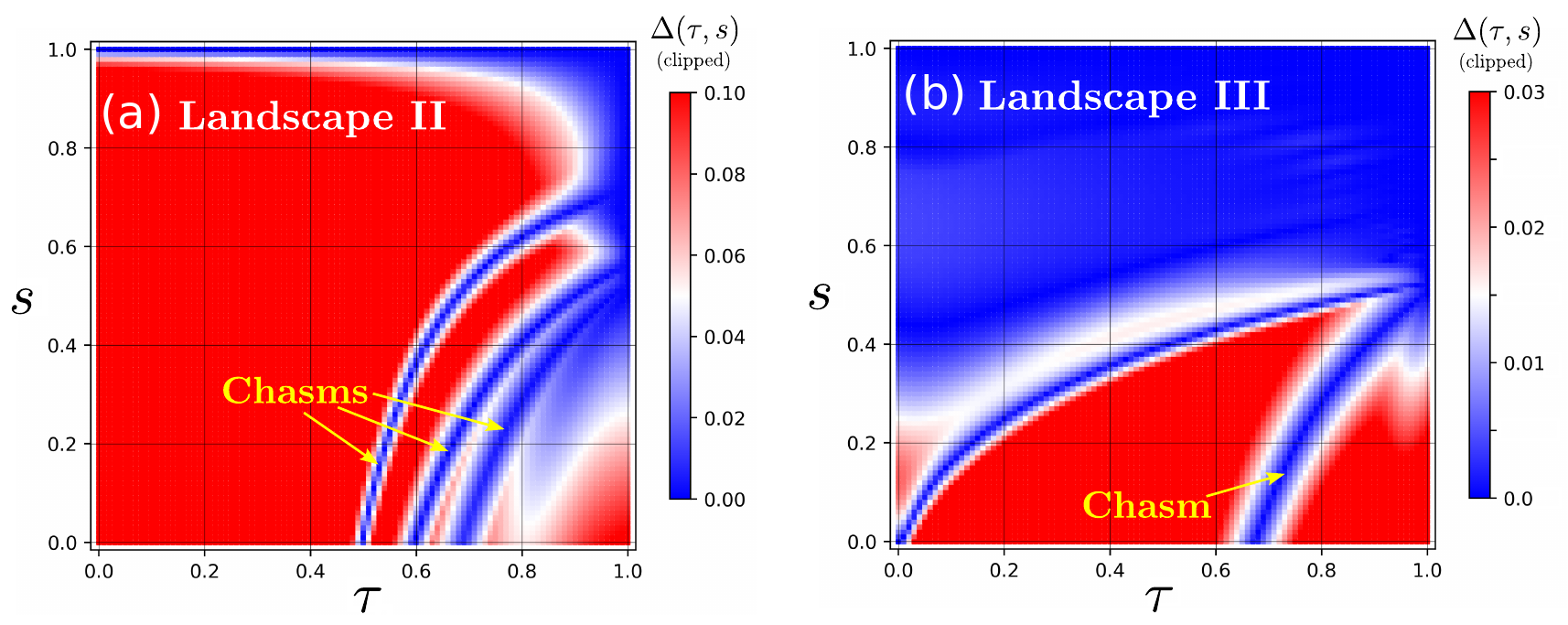}
\caption{Failure of the antiferromagnetic driver $+(J_x)^2$ to overcome small energy gaps in Landscapes II and III. (a) Gap landscape of $H^{(10)}$ [Eq. (\ref{eq.H(10)})] for $(p,q)=(12,0)$. (b) Of $H^{(11)}$ for $(6,6)$. In both cases, one sees `chasms' of gap closure on the gap landscape. One cannot go from $(\tau,s)=(1,0)$ to $(0,0)$ without crossing these chasms.}
\label{fig.gap landscape.under Jx2}
\end{center}
\end{figure*}

The discussions above focused on $U_x$ as the secondary driver. We also tried $T_x$ and $V_x$, and found that they are inappropriate because one cannot avoid gap closures on their gap landscapes. One might also inquire how the antiferromagnetic interaction $+(J_x)^2$ would fare here. As mentioned in the Introduction, this is the traditional secondary driver that one uses to overcome first-order phase transitions in QA. Let us combine it with $J_x$ and define the following Hamiltonian for Landscape II
\begin{equation}
H^{(10)}(\tau,s)=(1-s) \, H^{(6)}(\tau) + \frac{s}{\tilde{n}}\, \left(\frac{J_x}{\tilde{n}}\right)^2
\label{eq.H(10)}
\end{equation}
where $\tau,s \in [0,1]$. Similarly for Landscape III, define $H^{(11)}(\tau,s)$ according to Eq. (\ref{eq.H(10)}) with $H^{(7)}$ in place of $H^{(6)}$. Figure \ref{fig.gap landscape.under Jx2} shows the gap landscapes of $H^{(10)}$ and $H^{(11)}$ realized with $(p,q)=(12,0)$ and $(6,6)$, respectively. In both cases, we see `chasms' of zero gaps, making it impossible to anneal from $(\tau,s)=(1,0)$ to $(0,0)$ without encountering gap closures. These chasms persist with increase in multiplet dimension. Hence, the antiferromagnetic operator is not very useful as a secondary driver for annealing Landscapes II and III.

\section{Summary and discussions}
\label{sec.conclusions}

In this paper, we proposed a theoretical framework for quantum annealing based on $\mathfrak{su}(3)$ algebra. This is a generalization of the angular momentum algebra framework widely used in the study of QA of non-frustrated fully-connected systems. Similar to $\mathfrak{su}(2)$ where total angular momentum $J^2$ is conserved, in $\mathfrak{su}(3)$ the quadratic Casimir operator Eq. (\ref{eq.Casimir.C1.definition}) is invariant, and this allows us to work with irreducible representations of $\mathfrak{su}(3)$, thereby avoiding the exponentially large Hilbert spaces of full-blown spin glasses. 

Another important advantage of $\mathfrak{su}(3)$ is that rugged energy landscapes can be easily constructed using its Cartan subalgebra. We proposed Landscapes I to III and studied their quantum annealings numerically.  The geometries of these energy surfaces were designed with physical systems in mind, and are meant to be motifs representative of actual problems. We addressed the problem of energy gap closure and first-order transitions, which is an important difficulty currently faced by QA. By first visualizing the two-dimensional gap landscape $\Delta(\tau,s)$, we proposed annealing paths in $s$-$\tau$ space that circumvent gap closures and avoid first-order transitions. The effectiveness of the proposed paths were assessed using the residual energy $R(T)$, as well as the dependence of the energy gap and $R(T)$ on multiplet dimension. We showed that all three landscapes can be annealed successfully using the $\mathfrak{su}(3)$ framework.

We paid much attention to the time evolution of the wave function, visualizing the annealing dynamics in both configuration and multiplet spaces. The effectiveness of $\mathfrak{su}(3)$ drivers derives from their ability to transport the wave function nonlocally out of local minima, thereby avoiding being trapped by them. This nonlocal transport has an intuitive picture in multiplet space, where the wave function is moved obliquely along the multiplet diagram. We also annealed the landscapes using the familiar $J_x$ operator, on its own and in conjunction with $\mathfrak{su}(3)$ drivers. Overall, our results suggest that annealing with $\mathfrak{su}(3)$ drivers is generally more effective than annealing with $J_x$.

In our treatment, the 2-driver paths $s(\tau)$ were curated manually by hand. This is somewhat tedious and may be impractical in actual applications. It is also prone to subjectivity. In this respect, Durkin has demonstrated how annealing paths for 2-driver Hamiltonians can be obtained using the Dijkstra algorithm \cite{Durkin19}. Susa and Nishimori also proposed a variational approach for optimizing one parameter paths, which can be generalized to two parameters \cite{Susa21}. 

In QA studies, it is important to understand how a model behaves in the thermodynamic limit (scaling of energy gap, order of phase transitions etc.). In this work, we presented numerical results for multiplet dimension $d_m$ up to 276 for Landscapes I and II and 1728 for III. While it is computationally feasible to go to higher multiplets, analytical approaches based on path integral \cite{Seki12, Nishimori96, Koh18} or mean-field approximation \cite{Botet83, Durkin19} are more suitable for handling large system sizes. In this case, instead of the irreducible representation, it is better to adopt the many-body picture Eq. (\ref{eq.Gell-mann.aggregates}). For instance, in terms of Gell-Mann matrices, $T_3=\frac{1}{2}\sum_i \lambda_i^3$ \cite{Pfeifer03}. It would be interesting to analyze our $\mathfrak{su}(3)$ models along these lines in a future work.

Traditionally in QA, one places two requirements on the driver operator $H_{\mathrm{D}}$. Firstly, it should possess a simple ground state so that we can initialize the wave function easily. The second condition is that it should not commute with the problem Hamiltonian $H_{\mathrm{P}}$, i.e. $[H_{\mathrm{P}},H_{\mathrm{D}}]\ne 0$. This second condition is obviously too weak and in general will not lead to a successful annealing. To illustrate, consider the cubic case $(J_z)^3$ pointed out by Seki and Nishimori \cite{Seki12}. With the transverse field $J_x$ as driver, both commutators $[J_z^2,J_x]$ and $[J_z^3,J_x]$ are non-zero, but the former system yields a second-order transition whereas the latter gives a first-order one. We feel that one should tighten the second condition by introducing a measure on the degree of non-commutativity. This is motivated by our observation that the presence of nonlocal drivers, which have non-zero matrix elements far from the diagonal, seem to improve the chances of attaining the global minimum. Consider the measure
\begin{equation}
m(H_{\mathrm{D}})=||[H_{\mathrm{P}},H_{\mathrm{D}}]||_F
\label{eq.measure of non-commutativity of HD}
\end{equation}
where $||\cdot||_F$ denotes Frobenius norm. Equation (\ref{eq.measure of non-commutativity of HD}) is invariant under unitary transformation, and does not depend on the basis. Given a problem Hamiltonian $H_{\mathrm{P}}$, we may want to select $H_{\mathrm{D}}$ in such a way that $m(H_{\mathrm{D}})$ is maximized. More generally, the driver $H_{\mathrm{D}}$ would depend on a path $\gamma$ in parameter space, and we could demand that the integral of $m(H_{\mathrm{D}})$ along $\gamma$ be maximized. We think that this furnishes a more structured approach to choosing the most effective driver in QA. How this can be implemented in practice is a subject for future study.

\begin{acknowledgements}
The author thanks Prof. Shurtleff for helpful discussions.
\end{acknowledgements}

\appendix


\section{Labeling basis vectors in a SU(3) multiplet}
\label{app.sec.irrep of su(3)}

In Fig. \ref{fig.4 su3 potentials}, the label $i$ of the basis vectors in a $\mathfrak{su}(3)$ multiplet follows the prescription by Shurtleff. Here we briefly describe how $i$ maps onto the coordinates $(T_3,Y)$ on the multiplet diagram. For technical details, the reader is referred to the original papers \cite{Shurtleff23,Shurtleff24}.

\begin{figure}[h]
\begin{center}
\includegraphics[scale=0.45]{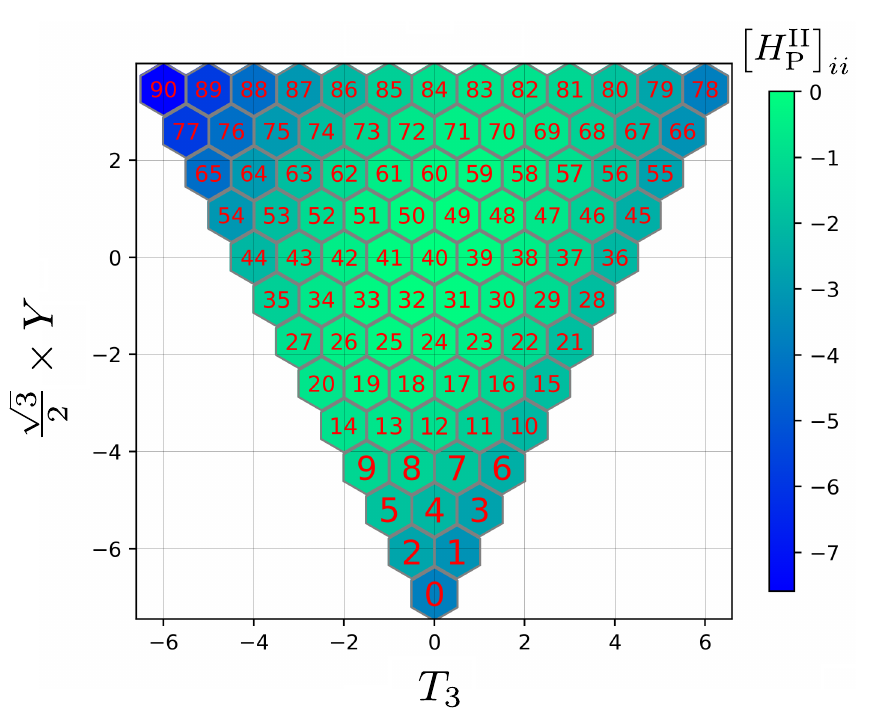}
\caption{Mapping the basis vectors of an irreducible representation onto the multiplet diagram, for the $(12,0)$-multiplet. Each hexagonal cell represents a state of the multiplet, and the red integer indicates the state index by Shurtleff. The color of a cell represents the value of the diagonal matrix elements $\left[H_{\mathrm{P}}^{\mathrm{II}}\right]_{ii}$, where $i$ is the state index of the cell.}
\label{fig.Landscape II.su(3) space}
\end{center}
\end{figure}

The basis of an irreducible representation is labeled from 0 to $d_m-1$. Figure \ref{fig.Landscape II.su(3) space} shows the labels of the multiplet $(p,q)=(12,0)$. The red integer in each hexagon indicates the state indices according to the prescription by Shurtleff. It increases from right to left horizontally, and then upwards vertically. The abscissas in Fig. \ref{fig.4 su3 potentials} refer to Shurtleff's state index.

\begin{figure}[h]
\begin{center}
\includegraphics[scale=0.5]{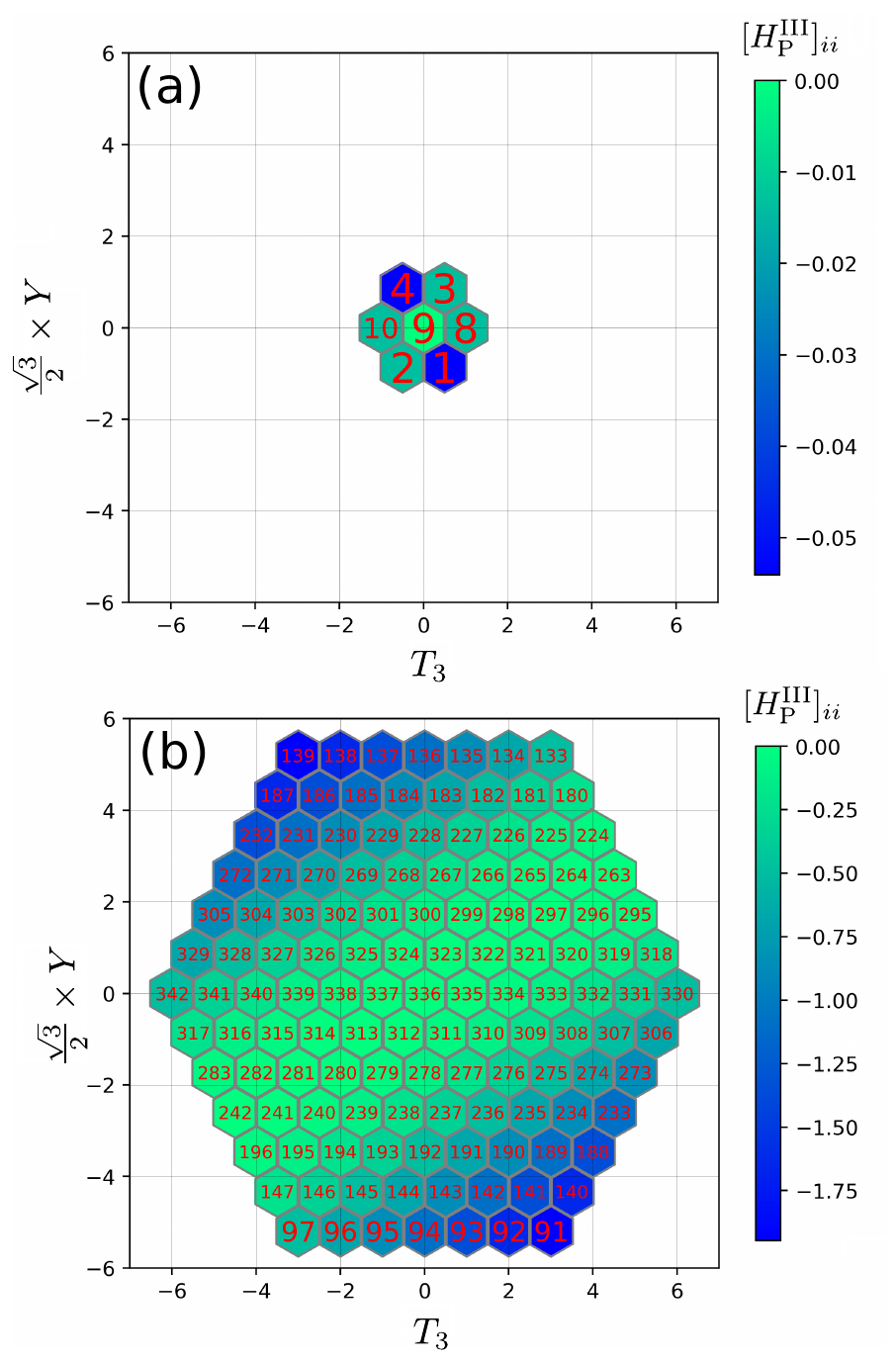}
\caption{Labeling of basis vectors for the $(6,6)$-multiplet, which has seven layers due to multiplicities. Panels (a) and (b) show Layers 2 and 7, respectively (see text). Red integers indicate state indices. Cell colors represent the diagonal matrix elements $\left[H_{\mathrm{P}}^{\mathrm{III}}\right]_{ii}$.}
\label{fig.Landscape III.su(3) multiplet space}
\end{center}
\end{figure}

Figure \ref{fig.Landscape III.su(3) multiplet space} shows the case of $(p,q)=(6,6)$. Due to multiplicities, the indexing is now more involved. The $(6,6)$-multiplet diagram can be imagined as the top-down view of a seven-layered hexagonal pyramid [see Fig. \ref{fig.su3 YvsT3 plot}(b)]. Each layer has one shell less than the layer below it. Let us call the top layer (peak of the pyramid) Layer 1, and the bottom layer (base of the pyramid) Layer 7. In Fig. \ref{fig.Landscape III.su(3) multiplet space}, panels (a) and (b) show Layers 2 and 7, respectively. Note that the state indices do not progress in an intuitive manner. For instance, in panel (a) the index progresses as 1, 2, 8, 9, 10, 3, 4. This is due to the way the states are being ordered in Shurtleff's algorithm.

%
%



\begin{thebibliography}{99}
\bibitem{footnote.su.convention} We use the notation $\mathfrak{su}(n)$ to denote Lie algebra.
\bibitem{footnote.01} Numerical integration was performed using the solve\_ivp function from the SciPy library, with `method' set to `DOP853', and both `rtol' and `atol' to `1e-11'.
\end{thebibliography}
\end{document}